\documentclass[screen]{acmart}

\AtBeginDocument{%
  }

\setcopyright{acmlicensed}
\copyrightyear{2018}
\acmYear{2018}
\acmDOI{10.1145/3705304}



\usepackage{tikz}
\usepackage{amsmath}
\usepackage{xspace}
\usepackage{enumitem}
\usepackage{subcaption}

\usepackage{multirow}
\usepackage{tabularx}
\usepackage{tablefootnote}
\usepackage{booktabs}

\usepackage{microtype}
\usepackage{wasysym}

\usepackage{amsthm}
\theoremstyle{acmdefinition}
\newtheorem{exmp}{Example}[section]

\usepackage{fancyhdr}

\usepackage{amssymb}
\usepackage{pifont}
\newcommand{\todo}[1]{\textcolor{black}{#1}}
\newcommand{\todel}[1]{{}}
\newcommand{\new}[1]{\textcolor{black}{#1}}
\newcommand{\tosem}[1]{\textcolor{black}{#1}}
\newcommand{\tosemrm}[1]{\textcolor{blue}{}}
\newcommand{\tool}{\textsc{Cerebro}\xspace}
\newcommand{\tosemrevision}[1]{\textcolor{black}{#1}}
\newcommand{\tosemrevisiontodo}[1]{\textcolor{black}{#1}}
\newcommand{\tosemminorrevision}[1]{\textcolor{black}{#1}}

\begin{document}


\title[Killing Two Birds with One Stone: Malicious Package Detection in NPM and PyPI using a Single Model of Malicious Behavior Sequence  \newline]{Killing Two Birds with One Stone: Malicious Package Detection in NPM and PyPI using a Single Model of Malicious Behavior Sequence}

\author{Junan Zhang}
\orcid{0009-0003-6485-9041}
\authornote{J. Zhang, Y. Huang, B. Chen, R. Wang, X. Peng are with the School of Computer Science and Shanghai Key Laboratory of Data Science, Fudan University, China.}
\authornote{J. Zhang and K. Huang contributed equally as first authors.}
\affiliation{%
\institution{Fudan University}
\city{Shanghai}
\country{China}
}

\author{Kaifeng Huang}
\orcid{0009-0000-1513-8254}
\authornote{K. Huang and B. Chen are the corresponding authors.}
\authornote{K. Huang is with the School of Software Engineering, Tongji University, China.}
\authornotemark[2]
\affiliation{%
\institution{Tongji University}
\city{Shanghai}
\country{China}
}

\author{Yiheng Huang}
\orcid{0009-0003-4722-3658}
\authornotemark[1]
\affiliation{%
\institution{Fudan University}
\city{Shanghai}
\country{China}
}

\author{Bihuan Chen}
\orcid{0000-0001-7238-7492}
\authornotemark[1]
\authornotemark[3]
\affiliation{%
\institution{Fudan University}
\city{Shanghai}
\country{China}
}

\author{Ruisi Wang}
\authornotemark[1]
\orcid{0009-0008-8404-4794}
\affiliation{%
\institution{Fudan University}
\city{Shanghai}
\country{China}
}

\author{Chong Wang}
\authornotemark[1]
\orcid{0000-0003-1424-6290}
\affiliation{%
\institution{Fudan University}
\city{Shanghai}
\country{China}
}

\author{Xin Peng}
\authornotemark[1]
\orcid{0000-0003-3376-2581}
\affiliation{%
\institution{Fudan University}
\city{Shanghai}
\country{China}
}

\renewcommand{\shortauthors}{Zhang et al.}

\begin{abstract}

\new{Open-source software (OSS) supply chain enlarges the attack surface \tosem{of a software system}, which makes package registries attractive targets for attacks. Recently, multiple package registries have received intensified attacks with malicious packages. Of those package registries, NPM and PyPI are two of the most severe victims. Existing malicious package detectors are developed with features from a list of packages of the same ecosystem and deployed within the same ecosystem exclusively, \tosemminorrevision{which is infeasible to utilize the knowledge of a new malicious NPM package detected recently to detect the new malicious package in PyPI.}
Moreover, existing detectors lack support to model malicious behavior of OSS packages in a sequential way.}

\new{To address the two limitations, we propose a single detection model using malicious behavior sequence, named \tool, to detect malicious packages in NPM and PyPI.~We curate a feature set based on a high-level abstraction of malicious behavior to enable multi-lingual knowledge fusing. We organize extracted features into a behavior sequence to model sequential malicious behavior. We fine-tune the \tosemrevision{pre-trained language model} to understand the semantics of malicious behavior. Extensive evaluation has demonstrated the effectiveness~of~\tool over the state-of-the-art as well as the practically acceptable efficiency. \tool has detected \todo{683} and \todo{799} new malicious packages in PyPI and NPM, and received \todo{707} thank letters from the official PyPI and NPM teams.}

\end{abstract}

\maketitle


\section{Introduction}


As the adoption of open source software (OSS) continues~to grow, the security concerns about OSS supply chain have attracted increased attention~\cite{oss_supply_chain_security, snykossreport, synopsysossreport, gu2022investigating, ladisa2022taxonomy}. The frequency and impact of OSS supply chain attacks have reached unprecedented levels. According to Sonatype~\cite{sonatype}, there has~been an astonishing average annual increase of 742\% in OSS supply chain attacks over the past three years. According to Gartner's prediction~\cite{gartnerpredict}, 45\% of organizations worldwide will~have experienced OSS supply chain attacks by 2025. This~alarming trend can be attributed to the expanded attack surface~through \tosemminorrevision{the} OSS supply chain. For example, installing an NPM package introduces an average of 79 transitively dependent packages and 39 maintainers which can be exploited; and some popular packages influence more than 100,000 packages and thus become attractive targets for attacks~\cite{zimmermann2019small}.

  
 



\textbf{Problem.} One of the OSS supply chain attacks is to inject malicious code into packages hosted in popular package~registries. Package registries NPM and PyPI have~been~flooded with malicious packages, as revealed by recent reports~\cite{npmpoisoning, pypipoisoning, sonatypereportcryptominers, pypiteammalwarereport, sonatypereport}. For example, the PyPI team removed over~12,000 packages in 2022, which were mostly malware~\cite{pypiteammalwarereport}; and the Sonatype team caught 422 malicious NPM packages~and 58 malicious PyPI packages in December 2022~\cite{sonatypereport}, mostly data exfiltration through typosquatting or dependency confusion attacks. An alarming incident occurred in December 2022~was the dependency confusion attack on PyTorch~\cite{pytorchcompromised}. The attack exploited the fact that the nightly built version of PyTorch depended on a package named ``torchtriton'' from the PyTorch nightly package index. However, a malicious package~with the same name was uploaded to PyPI. Since PyPI takes precedence over the PyTorch nightly package index, the malicious ``torchtriton'' was installed instead of the legitimate version. The prevalence of malicious packages in NPM and PyPI calls for practically~effective malicious package detection system.



\textbf{Existing Approaches.} Several approaches have been proposed to detect malicious packages in NPM and PyPI. They can be classified into rule-based~\cite{zahan2022weak, vu2020typosquatting, bandit4mal, Ossdetectbackdoor, malwarechecks, duan2020towards, taylor2020defending}, unsupervised learning~\cite{garrett2019detecting, Liang2021, ohm2020supporting}, and supervised learning approaches~\cite{fass2018jast, fass2019jstap, sejfia2022practical, ohm2022feasibility, liang2023needle, ladisa2023feasibility}. Rule-based approaches often rely on predefined rules about package metadata (e.g., package name) and suspicious imports and method calls. They often incur high false positives, which is far from reaching practical demands~\cite{vu2023bad}. Learning-based approaches capture malicious behavior as a set of discrete features. They overlook the sequential nature of malicious behavior which~is~usually composed of a sequence of suspicious activities, hindering~the practical effectiveness. Moreover, except for \cite{Ossdetectbackdoor, taylor2020defending, duan2020towards}, these approaches are designed and evaluated for one ecosystem (i.e., either NPM or PyPI). OSS Detect Backdoor~\cite{Ossdetectbackdoor} and Taylor et al.~\cite{taylor2020defending} adopt lightweight rules to support different~ecosystems, while Duan~et al.~\cite{duan2020towards} use heavyweight rules that require both static and dynamic program analysis.

\new{\textbf{Limitations.} We summarize two limitations that hinder the effectiveness of existing approaches. The first limitation~is that the knowledge \tosemminorrevision{of} malicious packages from different ecosystems is not sufficiently leveraged. 
Although there are evident clues \tosemminorrevision{of} attackers transcribing malicious packages and spreading \tosemminorrevision{them across} multiple ecosystems~\cite{multilanguageevidence}, \tosemminorrevision{there has been limited action in addressing this evolving threat landscape.}
Moreover, the NPM and PyPI teams do not publicly release the entire collection~of~malicious packages for preventing potential misuse or exploitation, and hence the publicly available datasets of malicious NPM and PyPI packages~\cite{ohm2020backstabber, duan2020towards} are small in size, which~are significantly smaller than the reported ones~\cite{sonatypereport, pypiteammalwarereport}. 
\tosemminorrevision{The malicious package detection may be less effective when the available malicious package data is limited.}
}
\new{The second limitation is that sequential knowledge is missing in existing approaches. Malicious packages typically conduct a sequence of suspicious activities to achieve an attack. However, existing rule-based and learning-based approaches fail to consider the sequential nature of malicious behavior. Consequently,~false positives and negatives may arise due to imprecise modeling.}

 


  

\todel{The second challenge is \textit{how to model malicious behavior in a sequential way such that maliciousness can be precisely captured.} Malicious packages typically conduct a sequence of suspicious activities to achieve an attack. However, existing rule-based and learning-based approaches fail to consider the sequential nature of malicious behavior. Consequently,~false positives and negatives may arise due to imprecise modeling.}

\new{\textbf{Our Approach.} To address these two limitations, the challenges become \textit{1) how to leverage the knowledge \tosemminorrevision{of} malicious packages from different ecosystems in a unified way such that multi-lingual malicious package detection can be feasible,} and \textit{2) how to model malicious behavior in a sequential way such that maliciousness can be precisely captured.} To that end, we propose a single model, named \tool, to detect~malicious packages in NPM and PyPI using malicious behavior sequences.} 

\new{\tool has three key components which are \tosemminorrevision{the} feature extractor, behavior sequence generator, and maliciousness classifier. The feature extractor employs static analysis to extract a set of \todo{16} features. We curate this feature set based on a high-level abstraction of malicious behavior, which is language independent. Thus, it enables multi-lingual knowledge fusing across NPM and PyPI, and solves the first challenge. Notably, \tool is focused solely on static features and does not consider metadata or dynamic features,~as it aims to strike a balance between effectiveness~and~efficiency.}
 
 
The behavior sequence generator produces a behavior sequence of a package by organizing extracted features based~on their likelihood of execution and their sequential order in execution, which solves the second challenge. We determine the execution likelihood according to the time phase of execution (i.e., install-time, import-time or run-time), and we determine the sequential order according to generated call graph. 

 
\tosemrevision{The maliciousness classifier employs a fine-tuned language model~\cite{kenton2019bert, raffel2020exploring, liu2019roberta} on the generated behavior sequence to determine whether a package is malicious or not.} We transform~the~behavior sequence into a textual description to ease the semantic understanding of malicious behavior. We fine-tune the model on NPM and/or PyPI packages into a binary classifier.



\textbf{Evaluation.} To evaluate the effectiveness and efficiency of \tool, we conduct extensive experiments on a dataset of \tosemrevision{2,675} malicious and \todo{7,391} benign NPM and PyPI packages. \tosemrevision{Our evaluation results have demonstrated that \tool outperforms the state-of-the-art by an average of 10.0\% in precision and 7.4\% in recall in the mono-lingual scenario (i.e., train and test~on~the same ecosystem), and by 9.9\% in precision and 8.9\% in recall in the bi-lingual scenario (i.e., train on two ecosystems and test on one of the two ecosystems).}~\tool takes~an average of 10.5 seconds to analyze a package, which~is~practically efficient. Further, we conduct an ablation~study~to~validate the contribution of behavior sequence.



\new{To evaluate the usefulness of \tool in practice, we~run \tool on the newly-published packages in PyPI and NPM over \todo{8} and \todo{7} months. From \todo{923,638} newly-published package versions, we detect \todo{5,976} potentially malicious package versions. After our manual confirmation, we detect \todo{683} malicious~PyPI package versions and \todo{799} malicious NPM package versions. We report these detected malicious package versions to the official PyPI and NPM teams. All these malicious package versions have been removed by the official teams. \todo{775}~of them have already been removed before our report, and hence we receive \todo{707} thank letters for the remaining ones.}


\textbf{Contributions.} Our work makes following contributions.
\begin{itemize}[leftmargin=*]
    \item We proposed and implemented a \todel{system}\new{single model} using malicious behavior sequence, named \tool, to detect malicious packages in NPM and PyPI.
    \item We conducted extensive experiments to demonstrate~the~effectiveness and efficiency of \tool.
    \item We detected \todo{683} and \todo{799} new malicious packages in PyPI and NPM, and received \todo{707} thank letters from the official team of PyPI and NPM.
\end{itemize}

\section{Threat Model}\label{threat}

\begin{figure}[!t]
    \centering
    \includegraphics[width=0.75\linewidth]{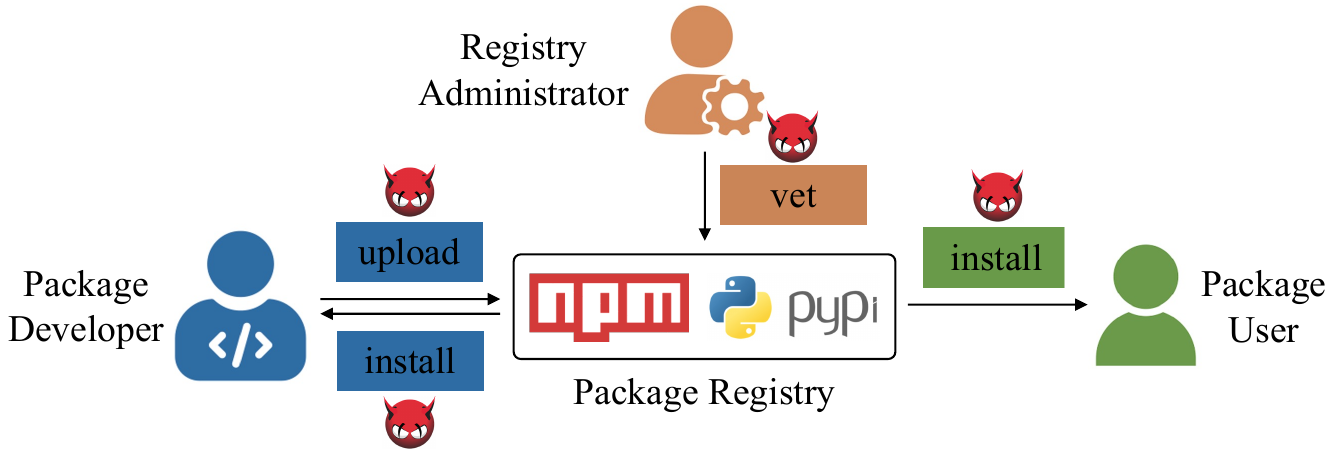}
    \vspace{-7pt}
    \caption{Threats in the Package Registry Ecosystem}\label{fig:threats}
\end{figure}

\tosem{We discuss the stakeholders and their activities involved in the development and distribution of OSS. Different from existing work~\cite{ladisa2022taxonomy}, our threat model \tosemminorrevision{is particularly} focused on the threats of using package registries, i.e., NPM~\cite{npmjs} and PyPI~\cite{pypi}.} NPM and PyPI are widely recognized as the primary package registries for hosting Python and JavaScript packages, respectively. To facilitate package management, PyPI is seamlessly integrated with the \textit{pip} package manager, which is the default package manager for Python. Similarly, NPM is closely integrated with the \textit{npm} package manager. As shown in Figure~\ref{fig:threats}, Package Developers (PDs) develop and maintain packages, and use the package manager to upload their packages to the package registry. Registry Administrators (RAs) then vet the uploaded packages and determine whether they should be published. Once published, packages become publicly available on the package registry. Package Users (PUs) then leverage the package manager to conveniently download and install desired packages for use in their own projects. We analyze the threats in these package management steps, which involve three key stakeholders (i.e., PDs, RAs and PUs).


\subsection{Threats in Developing Packages}

The threats in developing packages allow the injection~of~malicious code into packages. We summarize two key threats.

\textbf{Weak/Compromised Credentials.} Various tools (e.g.,~version control systems and~build~systems) are used during the development of packages. Hence, account hijacking may happen~due to weak credentials~\cite{Khandelwal2018}, or compromised credentials by exploiting vulnerabilities in these tools~\cite{Holmes2018, koishybayev2022characterizing}.~As~a~result, attackers can gain access to PDs' accounts, and hence have the privilege to~inject~malicious~code.

\textbf{Weak Governance in Collaborative Development.} Packages are collaboratively developed~by OSS community with many PDs. However, the governance of PDs~is weak.~First, malicious contributors can fool the development team of an existing package. They may first pretend~to~be~benign to~gain trust by committing~useful features and then secretly commit malicious code, or submit pull requests that fix bugs or add useful features but also include additional malicious code \cite{Gilbertson2018}. Second, benign PDs of an existing package may add malicious PDs into the team \cite{Sparling2018}~or~become malicious PDs~by~social engineering tactics, or the ownership is transferred to malicious PDs~\cite{Cimpanu2018}. As a result, malicious PDs have the full right~to~inject malicious code and publish malicious packages. Third, instead of infecting existing packages, attackers may first publish a benign and useful package, wait until it is used, and then update it to include malicious code~\cite{NPM2019}, or launch squatting attacks to inject malicious code \tosemminorrevision{into} a new package whose name is similar to a popular and benign package~\cite{vu2020typosquatting}.




\subsection{Threats in Vetting Packages} 

The threats in vetting packages allow malicious packages to be publicly available to PUs. We summarize two key threats.


\textbf{Ineffective/Insufficient Package Vetting.} A vetting system is often employed by RAs to first automatically identify suspicious packages and then manually triage them.~However, the vetting system lacks effectiveness, which allows malicious packages to evade detection and be published. This~is~evidenced by the recent report from Snyk~\cite{malware_snyk}, which~has~documented around 6,800 malicious packages on PyPI and~NPM~since~the beginning of 2023. One major contributing factor~is the overwhelming flood of suspicious packages~and~the~limited resources and budgets for RAs. For example, PyPI~had~one~person on-call to hold back weekend malware rush~\cite{pypi_one_person}. A recent interview with PyPI's RAs also confirms this dilemma~\cite{vu2023bad}. Consequently, the manual triage in the vetting system is not sufficient, allowing malicious packages to bypass the system. 


\textbf{Insecure Governance of RAs.} As package registries are often maintained by the OSS community, not all RAs can be blindly trusted. Attackers can disguise themselves~as~trustworthy developers to gain RAs' trust; and they may employ social engineering tactics to gain control of RAs' accounts. As a result, such malicious RAs have the right to grant and publish malicious packages to registries.

\subsection{Threats in Using Packages}\label{threats:invocation}

The threats in using packages allow malicious behaviors~to~be potentially triggered. We summarize two key threats.

\textbf{Weak Awareness of Security.} When PUs (including application developers and package developers) input the package name, the package manager searches and retrieves the corresponding package from the package registry. Unfortunately, this process is susceptible to various attacks. Attackers often exploit techniques such as typosquatting~\cite{taylor2020defending}, combosquatting~\cite{vu2020typosquatting} and dependency confusion~\cite{depconfusion} to deceive PUs into mistakenly installing malicious packages. Attackers may also employ search engine optimization (SEO) poisoning or phishing techniques~\cite{Kadouri2023} to advertise their packages on package registries, increasing the risk of unintended installation. 

\textbf{Weakness in Automated Dependency Resolution.} Package managers employ automated dependency resolution algorithms to select package versions when updating package~versions and installing transitive dependencies. As a result, malicious package versions can be silently installed. This places a long-standing burden on PUs, as they must consistently~monitor potential compromises among package updates~\cite{zimmermann2019small}.

\section{Preliminary and Motivation}\label{example}

We present preliminary information on triggering malicious behavior, and then show two motivating examples.

\subsection{Triggering Malicious Behavior}\label{example:trigger}

Once malicious code is present in an application's OSS supply chain (i.e., its direct and transitive dependencies) in NPM~and PyPI, malicious behavior can be triggered in three scenarios.

\textbf{Install-Time Execution.} Malicious code is often contained in install scripts that are automatically executed during package installation~\cite{ohm2020backstabber}. For PyPI packages, \textit{pip} automatically~executes the \texttt{setup.py} script present at the root of the package during installation. This script is responsible for performing any necessary preparation or configuration required for the installation. Similarly, NPM introduces a mechanism that~utilizes \textit{scripts} property in \tosemminorrevision{the} \texttt{package.json}~\cite{npmpackagejson}. This property allows developers to specify certain keys, such as \textit{preinstall} and \textit{postinstall}, to indicate the paths of scripts to be automatically executed. In this scenario, attackers inject the malicious payload into install scripts. The malicious payload is executed without requiring any additional action as long as the infected package is installed. This scenario provides attackers with the highest probability of successfully carrying out attacks. 

\textbf{Import-Time Execution.} Malicious code can also be executed when a package module is imported. This is achieved~in Python/JavaScript by injecting malicious code into the initialization file \texttt{\_\_init\_\_.py}/\texttt{index.js}, which is executed~by~default when~the~interpreter executes \textit{import}/\textit{require} statements. Further, the interpreter continues to load the imported module file and execute \tosem{the code at the global scope. The code at the global scope}~typically refers to statements that exist outside any method or class declaration. For the ease of presentation, we refer to \tosem{the code at the global scope} as the implicit ``main'' method. Attackers take advantage of this module import mechanism by injecting malicious code into this implicit ``main'' method. 

\textbf{Run-Time Execution.} Malicious code can also be executed at run-time during the normal control flow of applications. This is achieved by including the malicious code~in~a~legitimate method that is unlikely to raise suspicion while hoping that the infected method will be invoked by applications. This scenario provides the largest attack surface, as attackers can inject malicious code into any package method. However, it achieves the lowest probability of successfully carrying out attacks, as the infected method has a low chance of being called.

\begin{table*}[!t]
    \footnotesize
    \centering
    \caption{\tosemrevision{Feature Types used in Existing Literature}}\label{tab:feature_type}
    \vspace{-10pt}
    \begin{tabular}{m{2cm}m{6cm}m{6cm}}
        \toprule
    Feature Type & Description & Example Literature \\
    \midrule
    Metadata Feature & Package name, package size, uploader profile, etc & {Zimmermann et al. \cite{zimmermann2019small}}, {Taylor et al. \cite{taylor2020defending}}, {Vu et al. \cite{vu2020typosquatting}}, {Zahan et al. \cite{zahan2022weak}}  \\\hline
    Syntactic Feature & Elements and structures within the source code & {Liang et al. \cite{liang2023needle}}, {Vu et al. \cite{vu2021lastpymile}}, {Vu et al. \cite{vu2020towards}}, {Ladisa et al. \cite{ladisa2023feasibility}} \\\hline
    \tosemrevisiontodo{Semantic Feature} & Features that have explicit semantic meanings & \tosemrevisiontodo{Garrett et al. \cite{garrett2019detecting}}, {Fang et al. \cite{fang2021pbdt}} \\\hline
    Implicit Feature & Implicit features vectorized from program dependence graph, abstract syntax trees, etc & {Ohm et al. \cite{ohm2020supporting}} \\\hline
    Dynamic Feature &  Features captured during runtime execution &  {Duan et al. \cite{duan2020towards}}, {Ohm et al. \cite{ohm2020towards}} \\\hline
    \tosemrevisiontodo{Hybrid Feature} & Mixture of the single feature types & \tosemrevisiontodo{Liang et al. \cite{Liang2021}}, \tosemrevisiontodo{Duan et al. \cite{duan2020towards}}, \tosemrevisiontodo{{Sejfia et al. \cite{sejfia2022practical}}}, \tosemrevisiontodo{{Ohm et al. \cite{ohm2022feasibility}}} \\
    \bottomrule
\end{tabular}
\end{table*}

\begin{table*}[!t]
    \small
    \centering
    \caption{Feature Set to Model Malicious Behavior (\CIRCLE~= Supported, \Circle~= Unsupported)}\label{tab:behavior_set}
    \vspace{-10pt}
    \begin{tabular}{m{1.6cm}m{4.6cm}m{0.8cm}m{0.8cm}m{0.8cm}m{0.8cm}m{0.8cm}m{0.8cm}m{0.8cm}m{0.8cm}m{0.8cm}}
        \toprule
        Dimension & Feature Description & \rotatebox{75}{\tool} &  \rotatebox{75}{\textsc{Amalfi}\cite{sejfia2022practical}} &  \rotatebox{75}{\textsc{MalOSS}\cite{duan2020towards}} &  \rotatebox{75}{\textsc{PPD}\cite{Liang2021}} &  \rotatebox{75}{Garrett et al.\cite{garrett2019detecting}} &  \rotatebox{75}{Ohm et al.\cite{ohm2022feasibility}} & \rotatebox{75}{Fang et al.\cite{fang2021pbdt}} \\
        \midrule
        \multirow{6}{*}{Metadata} 
        & M1: suspicious package name & \Circle & \Circle & \CIRCLE & \CIRCLE & \Circle & \CIRCLE & \Circle \\
        & M2: suspicious maintainer & \Circle & \Circle & \CIRCLE  & \Circle & \Circle & \Circle & \Circle \\
        & M3: malicious dependencies & \Circle & \Circle & \CIRCLE & \Circle & \Circle & \Circle & \Circle \\
        & M4: abnormal publish time & \Circle & \CIRCLE & \CIRCLE  & \Circle & \Circle & \Circle & \Circle \\
        & M5: contain package install script & \Circle & \CIRCLE  & \CIRCLE  & \CIRCLE & \Circle & \CIRCLE & \Circle \\
        & M6: contain executable file & \Circle & \CIRCLE & \CIRCLE & \Circle & \Circle & \Circle & \Circle \\
        \midrule
        \multirow{5}{*}{ \shortstack[l]{Information \\ Reading}}
        & R1: import operating system module & \CIRCLE & \CIRCLE & \Circle & \Circle & \Circle & \Circle & \CIRCLE\\
        & R2: use operating system module call & \CIRCLE & \Circle & \CIRCLE  & \CIRCLE & \Circle & \Circle & \CIRCLE\\
        & R3: import file system module & \CIRCLE & \CIRCLE & \Circle  & \Circle & \Circle & \CIRCLE & \CIRCLE\\
        & R4: use file system module call & \CIRCLE & \Circle & \CIRCLE  & \CIRCLE & \CIRCLE & \Circle & \CIRCLE\\
        & R5: read sensitive information & \CIRCLE & \CIRCLE & \CIRCLE  & \CIRCLE & \Circle & \CIRCLE & \CIRCLE\\
        \midrule
        \multirow{3}{*}{\shortstack[l]{Data \\ Transmission}}
        & D1: import network module & \CIRCLE & \CIRCLE & \Circle  & \Circle & \Circle & \CIRCLE & \CIRCLE\\
        & D2: use network module call & \CIRCLE & \Circle & \CIRCLE  & \CIRCLE & \CIRCLE & \Circle & \CIRCLE\\
        & D3: use URL & \CIRCLE & \Circle & \CIRCLE  & \CIRCLE & \Circle & \CIRCLE & \CIRCLE\\
        \midrule
        \multirow{5}{*}{Encoding}
        & E1: import encoding module & \CIRCLE & \CIRCLE & \Circle & \Circle & \Circle & \Circle & \CIRCLE\\
        & E2: use encoding module call & \CIRCLE & \Circle & \Circle & \CIRCLE   & \Circle & \CIRCLE & \CIRCLE\\
        & E3: use base64 string & \CIRCLE & \Circle & \Circle  & \CIRCLE & \Circle & \CIRCLE & \Circle\\
        & E4: use long string & \CIRCLE & \Circle & \Circle  & \CIRCLE & \Circle & \Circle & \Circle\\
        \midrule
        \multirow{3}{*}{\shortstack[l]{Payload \\ Execution}}
        & P1: import process module & \CIRCLE & \CIRCLE & \Circle  & \Circle & \Circle & \CIRCLE & \CIRCLE\\
        & P2: use process module call & \CIRCLE & \Circle & \CIRCLE  & \Circle & \CIRCLE & \Circle & \CIRCLE\\
        & P3: use bash script & \CIRCLE & \Circle & \Circle  & \CIRCLE & \Circle & \Circle & \Circle\\
        & P4: evaluate code at run-time  & \CIRCLE & \CIRCLE & \CIRCLE & \Circle & \CIRCLE & \CIRCLE & \Circle\\
        \midrule
        Dynamic & /& \Circle & \Circle & \CIRCLE  & \Circle & \Circle & \Circle & \Circle\\
        \bottomrule
    \end{tabular}
\end{table*}

\subsection{\tosemrevision{Literature Survey on Malicious PyPI and NPM Package Features}}\label{example:features}

\tosemrevision{We conducted a comprehensive literature survey to understand malicious features in PyPI and NPM packages. We searched for related academic papers on Google Scholar using the keywords ``malicious'', ``PyPI'' and ``NPM''. We also employed a snowballing process to include referenced and cited papers.} \tosemrevision{In total, we obtained \tosemrevisiontodo{16} papers including long technical papers and short papers. We categorized the features used in those papers into six types, i.e., Metadata Feature, Syntactic Feature, Semantic Feature, Implicit Feature, Dynamic Feature, and Hybrid Feature, presented in Table \ref{tab:feature_type}. Additionally, we conducted a deeper comparison with the related literature focusing on Semantic and Hybrid Features. We excluded Metadata Features due to their lack of generalizability and adaptability to different package registries with varying metadata. Syntactic and Implicit Features were excluded because their results are often uninterpretable. Dynamic Features were also excluded due to their significant computational overhead and time-consuming nature.}

\tosemrevision{Based on above literature, we summarize the malicious features into six dimensions based on a high-level abstraction of malicious~behavior, i.e., metadata, information reading, data transmission, encoding, and payload execution, which is presented~in~Table~\ref{tab:behavior_set}. We can observe the existing approaches have varying implementation on the metadata and semantic features. Metadata, information reading and data transmission are the most frequently mentioned dimensions, followed by encoding and payload execution.}

\subsection{Motivating Examples}\label{example:example}

\todel{We use two examples to motivate the design of our approach.}

\begin{figure*}[!t]
    \centering
    \includegraphics[width=0.95\linewidth]{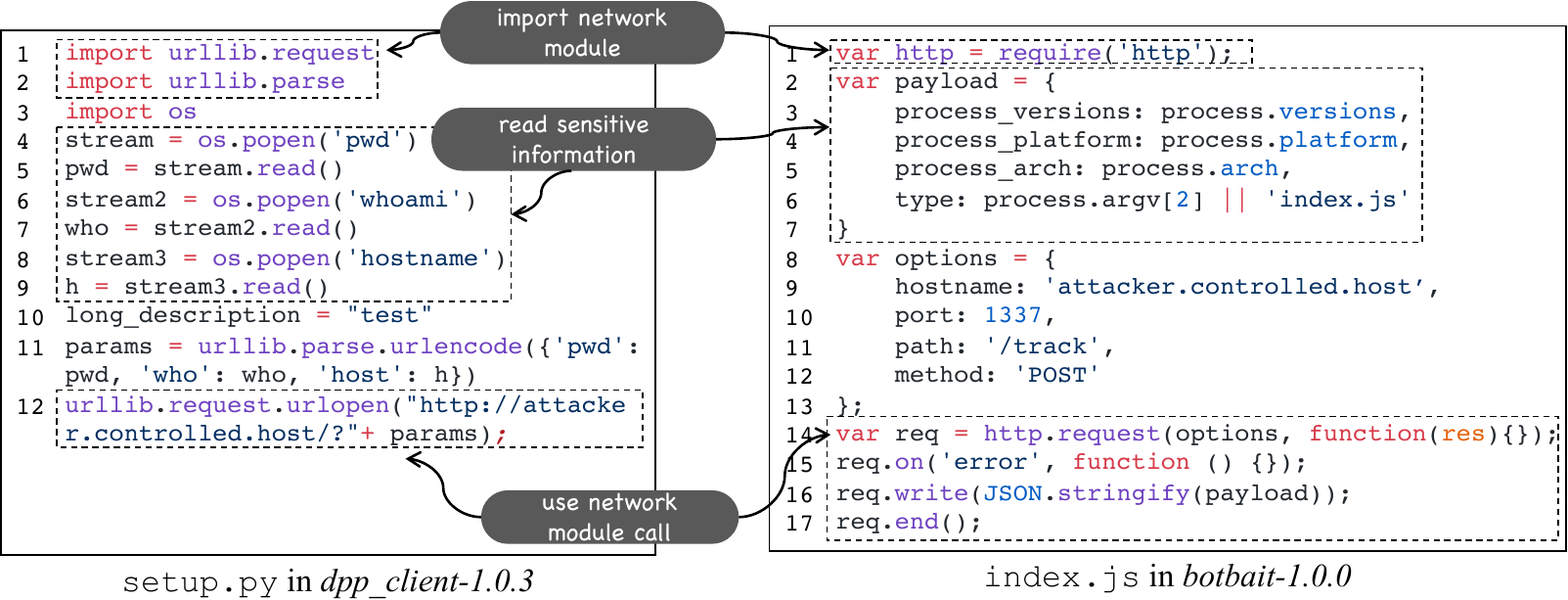}
    \vspace{-7pt}
    \caption{Malicious Packages from PyPI and NPM that Share Similar Malicious Behavior}\label{fig:example-cross-language}
\end{figure*}

\textbf{Example 1: Malicious Packages from PyPI and NPM Share Similar Malicious Behavior.} Figure~\ref{fig:example-cross-language} shows the~snippets of malicious PyPI package \textit{dpp\_client-1.0.3} and NPM package \textit{botbait-1.0.0}, which are taken from Backstabber's Knife Collection dataset~\cite{ohm2020backstabber}. The package \textit{dpp\_client-1.0.3} gathers sensitive information, including current working directory, user name and hostname, via executing CLI commands (e.g., \textit{pwd}, \textit{whoami} and \textit{hostname}) in \texttt{setup.py} (\tosemminorrevision{Lines}~4~to~9). Similarly, the package \textit{botbait-1.0.0} collects sensitive information about process's version, platform and architecture in \texttt{index.js} (\tosemminorrevision{Lines} 2 to 7). Both packages send the sensitive~information to remote servers using different libraries,~i.e.,~\textit{requests} in \texttt{setup.py} (Line 12) and \textit{http} in \texttt{index.js} (\tosemminorrevision{Lines} 14 to 17). Although being implemented in different languages with distinct syntax, these packages exhibit similar malicious behavior, involving the read of sensitive information and the calling of network operations. Sonatype also found that some malicious PyPI and NPM packages shared the same author, contained the identical malicious code, and used the same offending URL to fetch content from the remote~\cite{sonatypereportcryptominers}. 



Therefore, this opens up new opportunities for fusing the knowledge of malicious packages from different ecosystems at a high-level abstraction and for designing a unified detection system that supports different ecosystems. Such a unified detection system can partially address the challenge of limited dataset of malicious packages. It can also allow~us~to~identify unknown malicious packages that may have already been encountered in other package registries. 


\begin{figure*}[!t]
    \centering
    \includegraphics[width=0.95\linewidth]{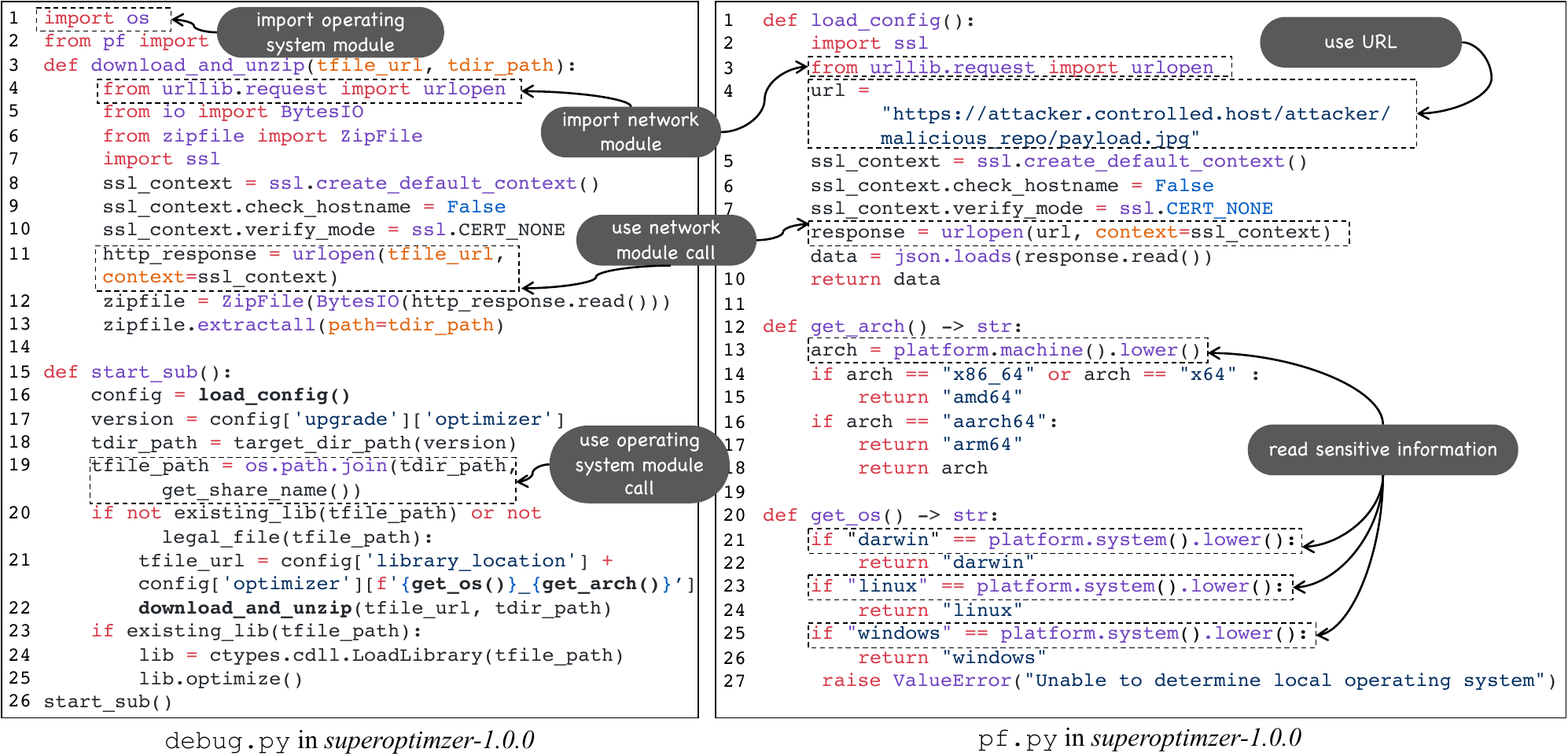}
    \vspace{-7pt}
    \caption{A Malicious Package from PyPI with a Malicious Behavior Sequence}\label{fig:example-sequence}
\end{figure*}

\textbf{Example 2: A Malicious Package Has a Malicious Behavior Sequence.} Figure~\ref{fig:example-sequence} shows the snippet of a malicious PyPI package \textit{superoptimzer-1.0.0}. The malicious behavior sequence is located in \texttt{debug.py} and \texttt{pf.py}, and is triggered upon the import of \texttt{debug.py}. When \texttt{debug.py} is imported, the method \texttt{start\_sub} is invoked in \texttt{debug.py} (Line 26).~In \texttt{start\_sub}, there are several inter-procedural calls~that carry out the attack, i.e., \texttt{load\_config} (Line 16), \texttt{get\_os} (Line 21), \texttt{get\_arch} (Line 21) and \texttt{download\_and\_unzip} (Line~22). In \texttt{load\_config} (\tosemminorrevision{Lines} 1 to 10 in \texttt{pf.py}), two suspicious activities, including using URL and using network call, download the payload (in a json format) from a remote server to \texttt{config}. Then, \texttt{config} is parsed to obtain \texttt{version}, \texttt{tdir\_path} and \texttt{tfile\_path} (\tosemminorrevision{Lines} 17 to 19 in \texttt{debug.py}). In \texttt{get\_os} (\tosemminorrevision{Lines} 20 to 27 in \texttt{pf.py}) and \texttt{get\_arch} (\tosemminorrevision{Lines} 12 to 18 in \texttt{pf.py}), sensitive information about the underlying operating system and architecture is read. Then, a target URL \texttt{tfile\_url}~is obtained by concatenating \texttt{config['library\_location']} and \texttt{config['optimizer']}  (Line 21 in \texttt{debug.py}). Finally, in \texttt{download\_and\_unzip} (\tosemminorrevision{Lines} 3 to 13 in \texttt{debug.py}), a network call is used to download the payload from \texttt{tfile\_url}. We can observe that the malicious behavior is often composed of a sequence of suspicious activities. 

Therefore, the modeling of malicious behavior sequence plays a crucial role in the accuracy~of malicious package~detection system. However, the state-of-the-art detection systems model malicious behavior as discrete features, and do not consider the sequential nature of malicious behavior.



\section{Methodology}\label{approach}

We first introduce the overview of our approach, then elaborate each step of our approach in detail, and finally present the implementation of our approach.

\subsection{Approach Overview}

\begin{figure*}[!t]
    \centering
    \includegraphics[width=0.9\linewidth]{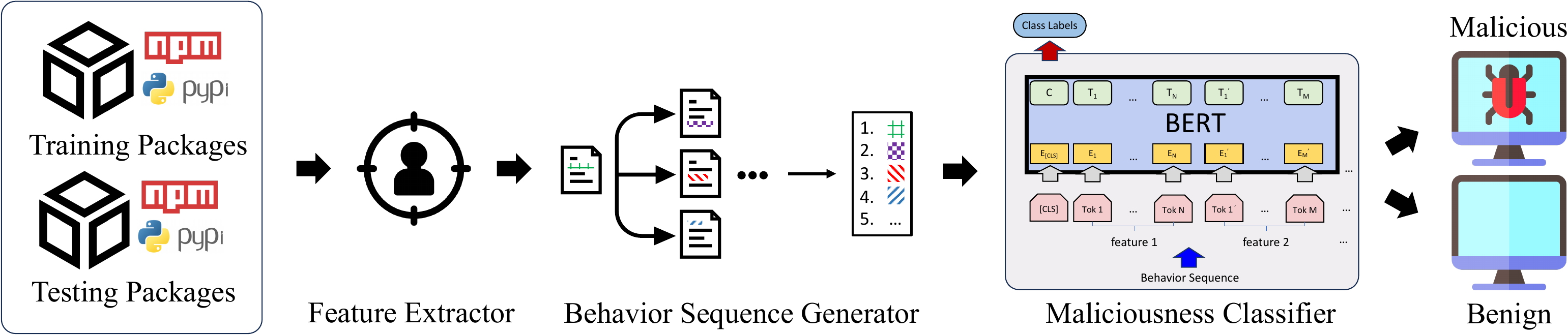}
    \vspace{-7pt}
    \caption{An Approach Overview of \tool for Detecting Malicious NPM and PyPI Packages}\label{fig:overview}
\end{figure*}

The goal~of~our approach is to support the analysis of NPM and PyPI packages with a unified model by leveraging the~bi-lingual knowledge \tosemminorrevision{of} malicious packages from NPM and PyPI as well as their~sequential malicious behavior knowledge. To achieve this, we propose \tool, a system designed~to detect malicious packages in NPM and PyPI ecosystems. The approach overview of \tool is shown in Figure~\ref{fig:overview}. Overall, it consists of three~key components, i.e., feature extractor (see Section \ref{approach:behavior:list}), behavior sequence generator (see Section \ref{approach:seq_gen}), and maliciousness~classifier~(see~Section~\ref{approach:maldetector}). 

In the offline training phase, \tool takes inputs~as~a~set of malicious and benign NPM and PyPI packages, uses feature extractor and behavior sequence generator to prepare~the~fine-tuning inputs, and fine-tunes a maliciousness classifier as the output. Similarly, in the online prediction phase, \tool takes an NPM or PyPI package as the input, uses feature extractor and behavior sequence generator to prepare the prediction input, and uses the maliciousness classifier to determine whether the package is malicious or benign.

Specifically, the feature extractor component extracts features from each source code file via static analysis. It enables bi-lingual knowledge fusing across NPM and PyPI through a set of language-independent features derived from a high-level abstraction of malicious behavior. Then, the behavior~sequence generator component generates behavior sequence for each package based on extracted features. It captures the~sequential relation among extracted features by call graph traversal. \tosemrevision{Finally, the maliciousness classifier component fine-tunes a pre-trained language model with behavior sequences into a binary classifier for malicious package detection.}


Our malicious package detection system addresses \tosemrm{all }the threats in Section~\ref{threat} as it trusts none of the stakeholders and prevents malicious packages from being published into package registries and thus from being used as well.

\subsection{Feature Extractor}\label{approach:behavior:list}

\tosemrevision{The details of our features are presented in~Table~\ref{tab:behavior_set}.} Each feature has a corresponding ID and a textual~description (e.g., R5: read sensitive information). We use common words to describe the feature to~make~use~of~the full capability of the language model in \tosemminorrevision{the} maliciousness classifier~(see Section \ref{approach:maldetector}). Due to space limitation, we compare our feature set with only state-of-the-art rule-based approach~\textsc{MalOSS}~\cite{duan2020towards}, unsupervised approaches Garrett et al.'s work~\cite{garrett2019detecting} and PPD~\cite{Liang2021}, and supervised approaches \textsc{Amalfi}~\cite{sejfia2022practical} and Ohm et al.'s work~\cite{ohm2022feasibility}. Notice that different approaches may have different implementations for a rule. For example, \textsc{MalOSS} considers a package name suspicious if it is similar to popular ones in the same registry, or if it is the same as popular packages in different registries, but with different~authors.~Differently, \textsc{PPD} determines a suspicious package name simply based on \tosemminorrevision{the} package~name's Levenshtein distance.

\textbf{Metadata.} The metadata includes package name (M1), maintainers (M2), dependencies (M3), publish time (M4),~and the existence of specific file types (M5 and M6). These metadata features only provide the abstraction for appearance characteristics without looking into code-level behavior. Besides,~analyzing metadata features relies on assumptions that~may~not hold in some circumstances. For example, information about popular packages may not exist universally in different package registries, or it is difficult to identify suspicious maintainers as prior knowledge. In order to make \tool~more~universally applicable, \tool does not leverage any metadata feature, but only uses features that can be extracted from~code. In this way, \tool operates independently of specific~registry characteristics and enhances its capability to detect malicious behavior across different package registry sources.

\textbf{Information Reading.} Attackers often attempt to read~sensitive information (R5), including personal information (e.g., account details, passwords, crypto wallets and credit card~information) as well as machine-related information (e.g., run-time environment~and machine name). They either steal such information, or use such information to carry out the attack. Besides, to effectively read sensitive information, attackers often leverage utility libraries provided by operating system~and file system. To detect such behavior, we focus on identifying the utilization of these libraries in the package,~including~module imports (R1 and R3) and method calls (R2 and R4). 

\textbf{Data Transmission. } Attackers often manipulate malicious packages to enable data transmission, either by downloading payloads or sending sensitive information. In order to detect such behavior, we consider the import of network modules (D1) and the use of network-related method calls~(D2)~as~suspicious indicators. Moreover, we identify string literals written in a URL format as an additional suspicious behavior~(D3). By examining the package code, we search for strings that adhere to the format of a URL. This indicates the presence of potential data transmission activities, such as communication with external servers or services.


\textbf{Encoding.} Attackers often leverage encoding methods~to obfuscate the malicious characteristics of their code, making it less noticeable and harder to detect. Therefore, we identify suspicious behavior related to the import of encoding modules (E1) and the presence of encoding calls within the code (E2). Besides, we also heuristically detect the use of base64 strings (E3) and the use of long strings (E4), which are known to be used in obfuscating or hiding malicious code~\cite{canali2011prophiler}.



\textbf{Payload Execution. } One of the goals of attackers~is~to~execute the payload downloaded to the malicious package. This suspicious behavior encompasses the import of process modules (P1) and the use of process-related function calls (P2). For example, the call to \texttt{Popen} in the \texttt{subprocess} module~in Python is considered as suspicious, as it allows the execution of arbitrary commands. The presence of such calls raises concerns about potential payload execution within the package. \tosem{It could either be benign or malicious in our context.} Further, we also identify suspicious behavior related to the use of bash scripts (P3). By analyzing the package code,~we match patterns of commands, including \textit{python <file>.py} and \textit{wget https://url}. The detection of these patterns suggests the potential execution of commands or scripts that may facilitate payload execution. Moreover, we identify run-time code evaluation (P4) such as \texttt{eval} as a suspicious behavior as it may execute the downloaded payload.

\textbf{Dynamic}. Dynamic features are collected by executing a package, which is heavyweight. Only \textsc{MalOSS}~\cite{duan2020towards} leverages these features, and we do not report the detailed list of dynamic features in Table~\ref{tab:behavior_set}. We do not consider dynamic features to strike a balance between effectiveness~and~efficiency.

The four abstraction dimensions, information reading, data transmission, encoding, and payload execution, are often used in combination to launch attacks. For example, attackers first read sensitive information and then send it through data transmission, or attackers first use data transmission to download payload, and then execute the payload. Encoding is leveraged to further hide the previous malicious behavior.

After introducing the feature set, we introduce how to extract these features. Generally, our feature extractor parses the abstract syntax tree (AST) of each source code file in~a~package to match patterns of features. Formally, the extracted~feature instances are denoted as $\mathcal{F}$. Each feature instance $f$ in $\mathcal{F}$~is~denoted as a 3-tuple $\langle m, l, id \rangle$, where $m$ and $l$ respectively denote the method and the line number where the feature instance is extracted, and $id$ denotes the ID of the feature. 

\begin{figure}[!t]
    \centering
    \includegraphics[width=0.8\textwidth]{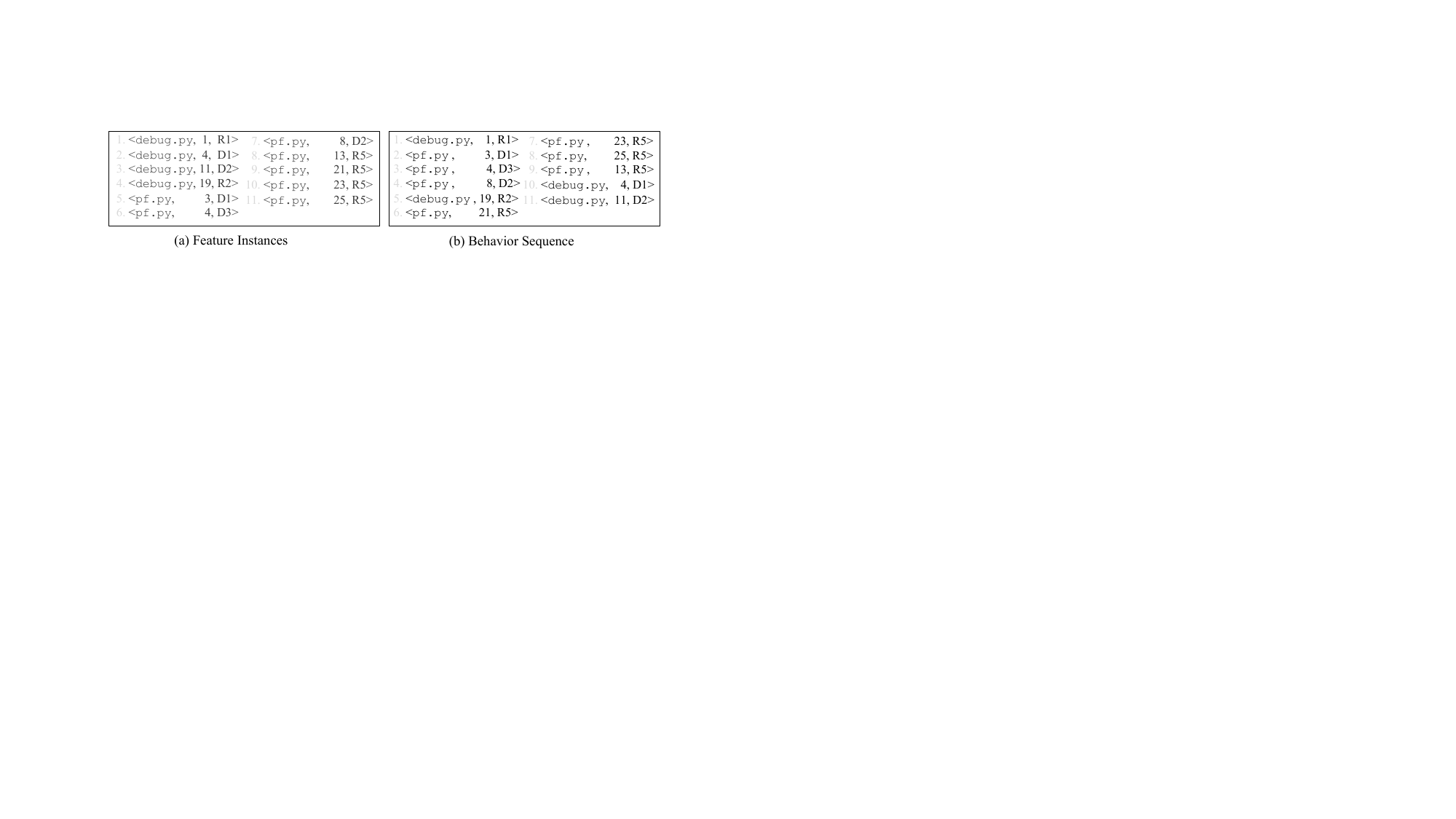}
    \vspace{-7pt}
    \caption{Extracted Feature Instances and Generated Behavior Sequence for the Package in Figure~\ref{fig:example-sequence}}\label{fig:example1}
\end{figure}

\begin{exmp}
Figure~\ref{fig:example1}(a) presents the extracted feature instances for the two source code files in Figure~\ref{fig:example-sequence}. For the ease of understanding, we use the source code file name instead of the method name in the extracted feature instances.
\end{exmp}

\subsection{Behavior Sequence Generator}\label{approach:seq_gen}

Our behavior sequence generator has three steps to organize the extracted feature instances into a behavior sequence based on their likelihood of execution and their sequential order in execution. The first step prioritizes the methods in a package according to the three triggering scenarios introduced~in~Section~\ref{example:trigger}, which is used to determine the execution likelihood of a feature instance. The second step constructs the call~graph of a package, which is used to determine the execution order of a feature instance. The last step generates the behavior~sequence of a package by querying sub call graphs in the order of prioritized methods and traversing them.

\textbf{Prioritizing Methods.} We first abstract the entire package into a collection of methods. The \tosem{code at the global scope} of each source code file is modeled into an implicit ``main'' method. Then,~we identify and prioritize the methods based on their triggering scenarios. As discussed in Section~\ref{example:trigger}, there are~three triggering scenarios, i.e., install-time, import-time and run-time~execution. Overall, methods that are executed at install-time~have high priority, followed by methods executed at import-time. Methods that are executed at run-time have low priority.

For methods executed at install-time, they are the implicit ``main'' methods of \texttt{setup.py} or script files specified in the \textit{scripts} property of \texttt{package.json} by \textit{preinstall}, \textit{install} and \textit{postinstall}. For methods executed at import-time, they consist of the implicit ``main'' methods in \texttt{\_\_init\_\_.py}, \texttt{index.js} and the imported module files. For methods executed at run-time, they encompass all publicly accessible methods, excluding private methods. In order to identify these methods, we need to exclude any method marked as private. In Python, a method is considered private when its name is prefixed with a double underscore (``\_\_''). In JavaScript, private methods are indicated by a hash (``\#'') prefix before the method name\cite{jshash}.



When this step finishes, we obtain a prioritized method~list, denoted as $\mathcal{M}$. Notice that the methods under the same triggering scenario are ranked in alphabet order by their names.

\textbf{Constructing Call Graph.} We employ static analysis techniques to~construct the call graph of a package, which serves as the foundation for subsequent analysis. Specifically, we denote the call graph $\mathcal{G}$ as a 2-tuple $\langle \mathcal{M}, \mathcal{E} \rangle$, where $\mathcal{M}$ denotes the prioritized list of methods in the previous step, and~$\mathcal{E}$~denotes the total set of calling edges. Each calling edge $e$ in $\mathcal{E}$ is denoted as a 3-tuple $\langle m_a, l, m_b\rangle$, which means that there is a method call at line $l$ in method $m_a$ that calls method $m_b$.


\textbf{Generating Behavior Sequence.} We iterate each method $r$ in $\mathcal{M}$ (i.e., in the order of execution likelihood), and query~$\mathcal{G}$ to get a sub call graph with its root being method~$r$.~We start from visiting $r$, and traverse the sub call graph in the~order of execution to generate the behavior sequence.

During the sub call graph traversal, when a method~$m$~is~visited, we retrieve the feature instances extracted from $m$ (denoted as $\mathcal{F}_m$) and obtain the calling edges starting from $m$~(denoted as $\mathcal{E}_m$). Then, we enqueue $\mathcal{F}_m$ and $\mathcal{E}_m$ together into a queue in the order of their line numbers. In other words, given $f \in \mathcal{F}_m$ and $e \in \mathcal{E}_m$, we compare $e.l$ and $f.l$ to determine their execution order. Next, we dequeue elements from the queue. If the dequeued element is a feature instance $f$, we append~$f$ into the behavior sequence $\mathcal{S}$; and if the dequeued element is a calling edge $e$, we start to recursively visit the called method $e.m_{b}$. This makes \tool to preserve the sequential information through inter-procedural analysis, ensuring that the generated behavior sequence reflects the~execution~order.


\begin{exmp}
Figure~\ref{fig:example1}(b) presents the generated behavior sequence when the implicit ``main'' method of \texttt{debug.py}~is visited. Notice that this implicit ``main'' method is executed when \texttt{debug.py} is imported. We can see that the extracted feature instances in Figure~\ref{fig:example1}(a) are organized according to their execution order in the two source code files in Figure~\ref{fig:example-sequence}.
\end{exmp}

\subsection{Maliciousness Classifier}\label{approach:maldetector}

\tosemrevision{The maliciousness classifier leverages pre-trained language models to have a semantic representation of the behavior sequence from a package and then classify it as either malicious or benign. We adopt BERT~\cite{kenton2019bert}, RoBERTa~\cite{liu2019roberta} and the encoder in T5~\cite{raffel2020exploring} as they are widely-used. Specifically, we first transform $\mathcal{S}$ into a textual description $\mathcal{D}$ to facilitate the bi-lingual knowledge fusing as well as to ease the semantic understanding of behavior sequence. Then, by feeding textual descriptions of malicious and benign packages, we fine-tune the pre-trained language model into a binary classifier that fits for our task of malicious package detection in NPM and PyPI.}




\textbf{Transforming Behavior Sequence into Textual Description.} We iterate each feature instance $f$ in $\mathcal{S}$, and append~the corresponding feature description (as shown in Table~\ref{tab:behavior_set}) into the textual description $\mathcal{D}$. When transforming the part of the behavior sequence that is generated from a method, we insert two short descriptions, namely \textit{start entry <file\_name>}~and \textit{end of entry}, to mark the beginning and end of the feature instances originating from the same method.~Consequently, we obtain a textual description for a package where the sequence of feature descriptions are concatenated through commas. As the \tosemrevision{pre-trained language model} is exposed to massive textual data during pre-training, it is enabled to understand the semantics of certain security-related terms (e.g., ``network'', ``URL'' and ``base64''). Hence, by using common words, the maliciousness classifier can benefit from the knowledge acquired by \tosemrevision{the pre-trained language model}.



\begin{figure}[!t]
    \centering
    \includegraphics[width=0.95\textwidth]{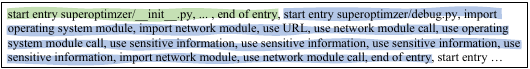}
    \vspace{-7pt}
    \caption{Textual Description of the Behavior Sequence Generated for the Package in Figure~\ref{fig:example-sequence}}\label{fig:example-behavior-text}
\end{figure}

\begin{exmp}
Figure \ref{fig:example-behavior-text} depicts the textual description of the behavior sequence of the package in Figure~\ref{fig:example-sequence}. To differentiate the descriptions generated from different methods,~we~highlight them with different colors. The green part~corresponds to the behavior executed in the implicit ``main'' method in \texttt{superoptimizer/\_\_init\_\_.py} at import-time. The blue part corresponds to the behavior executed in the implicit ``main'' method in \texttt{superoptimizer/debug.py} when \texttt{debug.py} is imported, which is the textual description of the behavior sequence in Figure~\ref{fig:example1}(b).
\end{exmp}

\textbf{Fine-Tuning \tosemrevision{Pre-Trained Language Models}.}  We append a fully connected layer and a softmax layer to \tosemrevision{each pre-trained language model's} architecture, and adopt the cross-entropy loss function to fine-tune \tosemrevision{the models into binary classifiers}. We feed the model with textual descriptions of both malicious and benign packages from PyPI and NPM, enabling the classifier to learn bi-lingual knowledge. The position embedding module and self-attention mechanism in the \tosemrevision{pre-trained language model} allow it to capture the sequential behavior knowledge.

\subsection{Implementation}

We have implemented \tool in \todo{3.7K} lines of Python code. To extract features, we employ tree-sitter~\cite{treesitter} to transform Python and JavaScript code into an AST representation, and identify suspicious behavior by matching the syntactic structures using AST queries provided by tree-sitter. To prioritize methods, we also use tree-sitter to parse each source code. To generate call graph, we leverage PyCG~\cite{salis2021pycg} for Python and Jelly~\cite{jelly, nielsen2021modular, moller2020detecting, feldthaus2013efficient} for JavaScript. During the fine-tuning process, we employ the Adam optimizer with a learning rate of 1e-6 and a batch size of 1. The model is trained for~3~epochs to ensure optimal learning and performance. \tosemrm{To comply with the token limit in BERT, we keep the first 512 tokens if the generated textural description~exceeds~512~tokens.}


\section{Evaluation}\label{sec:evaluation}

We first present the evaluation setup, and then report the evaluation results~of~research~questions.

\subsection{Evaluation Setup}

\todel{We conduct extensive experiments to evaluate the effectiveness, efficiency and usefulness of \tool.}

\textbf{Research Questions.} We design our evaluation to answer the following four research questions.

\begin{itemize}[leftmargin=*]
    \item \textbf{RQ1 Effectiveness Evaluation:} How is the effectiveness~of \tool, compared with the state-of-the-art malicious package detection approaches?
    \item \textbf{RQ2 Efficiency Evaluation:} How is the performance overhead of \tool?
    \item \tosemrevision{\textbf{RQ3 Dataset Scale Evaluation:} How does the scale of the dataset affect the effectiveness of \tool?}
    \item \textbf{RQ4 Ablation Study:} How behavior sequence contributes to the effectiveness of \tool?
    \item \textbf{RQ5 Usefulness Evaluation:} How useful is \tool in real-world detection on PyPI and~NPM?
\end{itemize}

\textbf{Dataset Collection.} We constructed the dataset as follows. For malicious packages, we collected malicious samples from three sources. We collected \new{438} malicious PyPI packages and \new{1,788} malicious NPM packages from Backstabber's Knife Collection~\cite{ohm2020backstabber}. We added \new{88} malicious PyPI packages from \textsc{MalOSS}'s dataset~\cite{duan2020towards}. \tosemrevision{We also sampled \new{361} packages out of a total of \tosemrevision{5,874} packages from pypi\_malregistry \cite{guo2023empirical}, with a confidence level of 95\% and a margin of error of 5\%.} For benign packages, we collected \new{2,398} benign PyPI packages from Vu et al.'s dataset~\cite{vu2023bad}. Following the procedure in prior work~\cite{ohm2022feasibility, vu2023bad}, we selected the \new{5,000} most depended upon packages in the NPM registry as the dataset of benign NPM packages. However, we successfully downloaded \new{4,993} of them. In total,~we curated a dataset of PyPI and NPM packages with \new{2,314} malicious packages and \new{7,391} benign packages, as listed in Table~\ref{table:dataset} where \textit{Mixed} denotes the summation across PyPI and NPM. \tosemrevision{We also provide the average file count and average thousand lines of code (KLOC) for each package.}

\begin{table}[!t]
    \centering
        \small
    \caption{\tosemrevision{Statistics of the Dataset}}\label{table:dataset}
    \vspace{-10pt}
    \begin{tabular}{m{2cm}m{1.6cm}m{1.6cm}m{1.6cm}m{1.6cm}m{1.6cm}m{1.6cm}}
        \toprule
        \multirow{2}{*}{Package Registry} & \multicolumn{3}{c}{Malicious} & \multicolumn{3}{c}{Benign} \\
        \cmidrule(lr){2-4}
        \cmidrule(lr){5-7}
        & \# Packages & \tosemrevision{Aver. Files \#} & \tosemrevision{Aver. KLOC \#} & \# Packages & \tosemrevision{Aver. Files \#} & \tosemrevision{Aver. KLOC \#} \\
        \midrule
        PyPI & 887 & 5.6 & 0.987 & 2,398 & 75.3 & 19.492 \\
        NPM & 1,788 & 4.4 & 1.642 & 4,993 & 214.3 & 9.256 \\
        Mixed & 2,675 & 4.8 & 1.425 & 7,391 & 53.4 & 12.577 \\
        \bottomrule
    \end{tabular}
\end{table}


\textbf{RQ Setup.} For \textbf{RQ1}, we aim to compare the effectiveness of \tool with state-of-the-art approaches. \tosemrevision{Specifically, we selected \textsc{Amalfi}~\cite{sejfia2022practical}, \textsc{SAP}~\cite{ladisa2023feasibility} and \textsc{MPHunter}~\cite{liang2023needle} as the state-of-the-art learning-based approaches}, and OSS Detect Backdoor~\cite{Ossdetectbackdoor} and Bandit4Mal~\cite{bandit4mal} as the state-of-the-art rule-based approaches. We adopted the same configuration used in the original paper~\cite{sejfia2022practical, ladisa2023feasibility, liang2023needle}. For OSS Detect Backdoor and Bandit4Mal, we set the threshold of 3 alerts to distinguish malicious and benign packages, which was observed as the optimal threshold by Vu et al.~\cite{vu2023bad}.~We did not compare with \textsc{MalOSS}~\cite{duan2020towards} as it used dynamic features and thus failed to run on many packages. We also did~not compare with Malware Checks~\cite{malwarechecks} as the PyPI team removed it recently. We split the dataset into training and testing dataset by 9:1. We measured the precision and recall of all the approaches with 10-fold cross-validation on the testing datasets. We compared these approaches in three detection scenarios, i.e., mono-lingual scenario (i.e., train and test on the same ecosystem), cross-lingual scenario (i.e., train on one ecosystem and test on the other), and bi-lingual scenario (i.e., train on two ecosystems and test on one of the two ecosystems). \tosemrevision{We trained \tool with BERT~\cite{kenton2019bert}, RoBERTa~\cite{liu2019roberta} and T5~\cite{raffel2020exploring} on the PyPI, NPM and Mixed training datasets, denoted by \tool$_{BERT}$, \tool$_{RoBERTa}$ and \tool$_{T5}$. We also trained \textsc{Amalfi} with decision tree (\textsc{Amalfi}$_{DT}$), naive bayes (\textsc{Amalfi}$_{NB}$) and SVM (\textsc{Amalfi}$_{SVM}$) on the same training datasets. We only evaluate \textsc{SAP} in both mono-lingual and bi-lingual scenarios, and \textsc{MPHunter} in the mono-lingual scenario for PyPI packages, as their do not support adaptation to other scenarios.}

For \textbf{RQ2}, we measured the time overhead of \tool~in offline training and online prediction. \tosemrevision{For \textbf{RQ3}, we evaluated the effectiveness of \tool across different dataset scales. We incrementally increased the proportion of our training dataset from 10\% to 100\% in 10\% steps. We trained \tool and \tosemrevision{\textsc{Amalfi}$_{DT}$} in the bi-lingual scenario using these varying dataset proportions. The trained models were then tested using the same testing dataset across all scales. The \tosemrevision{average F1-Score over 10-fold} cross-validation was used to measure overall effectiveness.}

For \textbf{RQ4}, we created three ablated versions of \tool and compared their effectiveness with the original version of \tool using the same dataset in \textbf{RQ1}. Specifically, to evaluate~the~impact of sequential behavior, we created “\tool w/o Seq” by removing behavior sequence generator and feeding textual descriptions directly into maliciousness classifier. To evaluate the impact of textual description transformation, we created ``\tool w/o Text'' by removing textual description transformation in maliciousness classifier. To evaluate the impact of \tosemrevision{pre-trained language models}, we created ``\tool w/ DT'' by removing behavior sequence generator and feeding features as a vector into~the~decision~tree.

For \textbf{RQ5}, we ran \tosemrevision{\tool using the BERT model} against the newly published packages in PyPI and NPM for \todel{over \todo{6} weeks}\new{over nine months}.  \new{This monitoring process started \tosemminorrevision{on} March 03 2023 for PyPI and April 01 2023 for NPM, and ended at October~30~2023}\todel{May~27~2023}. \tool analyzed \new{599,493} PyPI package versions and \new{324,145} NPM package versions. We manually confirmed the potentially malicious package versions flagged~by~\tool.\tosemrm{, and reported them to the official PyPI and NPM teams.} \tosem{Specifically, two authors manually assessed the maliciousness of the packages in the dimension of static characteristics, dynamic behaviors and third-party detection tools. Two authors conducted a manual assessment to determine the maliciousness of the packages. This process covered comprehending the source code, inspecting the dynamic behaviors, and referencing detection results from third-party detection tools (e.g., virustotal~\cite{virustotal}). The two authors are postgraduate and undergraduate students with security/malware background. The human inspection was conducted daily, costing one hour each day, over a period of ten months. The average number of reviewed packages daily was 30. All packages confirmed by two authors as malicious were reported to the official PyPI and NPM teams, all of which were accepted by the two registries. }Further,~by adding the confirmed~malicious and benign packages into the original training datasets (i.e., incremental learning),~we~retrained \tosemrevision{\tool using the BERT model} to evaluate whether \tool could learn from new data to improve its effectiveness. We also analyzed the malicious intentions and triggering scenarios of the new malicious packages.

\begin{equation}\label{eq:pre_rec}
    \begin{small}
    \begin{aligned}
        \vspace{-5pt}
        Pre. = \frac{\mid  MP_{tool} \cap MP_{test} \mid}{\mid MP_{tool} \mid }\\
        Rec. =  \frac{\mid  MP_{tool} \cap MP_{test}  \mid }{\mid MP_{test} \mid}\\
        \tosemrevision{F1-Score = \frac{2 \times Pre. \times Rec.}{Pre. + Rec.}}
    \end{aligned}
    \end{small}
\end{equation}

\tosem{\textbf{Evaluation Metric.} The precision (Pre.), recall (Rec.) used in \textbf{RQ1}, \textbf{RQ4}, \textbf{RQ5} and \tosemrevision{F1-Score used in \textbf{RQ3}} are defined in Equation~\ref{eq:pre_rec}. Specifically, $MP_{tool}$ denotes the set of malicious packages predicted by tools (e.g., \tool), and $MP_{test}$ denotes the set of true malicious packages in the testing dataset. Precision is calculated by the proportion of true positives (i.e., $MP_{tool} \cap MP_{test}$) in the predicted malicious packages (i.e., $MP_{tool}$), and recall is calculated by the proportion of true positives (i.e., $MP_{tool} \cap MP_{test}$) in the ground truth (i.e., $MP_{test}$).}

\begin{table}[!t]
    \centering
    \footnotesize
    \caption{\tosemrevision{Evaluation Result in Mono-Lingual Scenario for Learning-based Approaches}}\label{table:result-mono-lingual}
    \vspace{-10pt}
    \begin{tabular}{m{0.40cm}m{0.40cm}m{0.46cm}m{0.46cm}m{0.46cm}m{0.46cm}m{0.46cm}m{0.46cm}m{0.46cm}m{0.46cm}m{0.46cm}m{0.46cm}m{0.51cm}m{0.51cm}m{0.46cm}m{0.46cm}m{0.46cm}m{0.46cm}}   
        \noalign{\hrule height 1pt}
        \multirow{2}{*}{Train}  & \multirow{2}{*}{Test} & \multicolumn{2}{c}{\shortstack[l]{\tool$_{BERT}$}} & \multicolumn{2}{c}{\tosemrevision{\shortstack[l]{\tool$_{RoBERTa}$}}} & \multicolumn{2}{c}{\tosemrevision{\shortstack[l]{\tool$_{T5}$}}}  & \multicolumn{2}{c}{\shortstack[l]{\textsc{Amalfi}$_{DT}$}} & \multicolumn{2}{c}{\shortstack[l]{\textsc{Amalfi}$_{NB}$}} & \multicolumn{2}{c}{\shortstack[l]{\textsc{Amalfi}$_{SVM}$}} & \multicolumn{2}{c}{\tosemrevision{\textsc{SAP}}} & \multicolumn{2}{c}{\tosemrevision{\textsc{MPHunter}}}\\
    \cmidrule(lr){3-4}
    \cmidrule(lr){5-6}
    \cmidrule(lr){7-8}
    \cmidrule(lr){9-10}
    \cmidrule(lr){11-12}
    \cmidrule(lr){13-14}
    \cmidrule(lr){15-16}
    \cmidrule(lr){17-18}
    & & Pre. & Rec. & Pre. & Rec. & Pre. & Rec. & Pre. & Rec. & Pre. & Rec. & Pre. & Rec. & Pre. & Rec. & Pre. & Rec.\\ 
    \noalign{\hrule height 1pt}
    PyPI & PyPI & 95.8\% & 89.4\% & \tosemrevision{96.0\%} & \tosemrevision{91.7\%} & \tosemrevision{93.4\%} & \tosemrevision{70.9\%} & 81.9\% & 83.9\% & 59.5\% & 7.5\% & 30.3\% & 30.9\% & \tosemrevision{83.9\%} & \tosemrevision{61.5\%} & \tosemrevision{21.8\%} & \tosemrevision{85.4\%}\\
    NPM & NPM & 98.2\% & 91.8\% & \tosemrevision{98.5\%} & \tosemrevision{92.9\%} & \tosemrevision{98.9\%} & \tosemrevision{91.0\%} & 92.7\% & 86.0\% & 92.1\% & 77.1\% & 10.2\% & 26.7\%  & \tosemrevision{90.5\%} & \tosemrevision{35.4\%} & \tosemrevision{--} & \tosemrevision{--}\\
    \noalign{\hrule height 1pt}
    \end{tabular}
\end{table}

\begin{table}[!t]
    \centering
    \footnotesize
    \caption{Evaluation Result in Mono-Lingual Scenario for Rule-Based Approaches}\label{table:result-rule-based}
    \vspace{-10pt}
    \begin{tabular}{ccccc}
        \noalign{\hrule height 1pt}
        \multirow{2}{*}{Test} & \multicolumn{2}{c}{OSS Detect Backdoor} & \multicolumn{2}{c}{Bandit4Mal} \\
        \cmidrule(lr){2-3}
        \cmidrule(lr){4-5}
         & Pre. & Rec. & Pre. & Rec. \\ 
        \noalign{\hrule height 1pt}
        PyPI & 19.9\% & 60.4\% & 24.9\% & 74.4\%\\
        NPM & 41.1\% & 85.1\%  &  -- & -- \\
        \noalign{\hrule height 1pt}
    \end{tabular}
\end{table}


\subsection{Effectiveness Evaluation (RQ1)}

\todel{We report the effectiveness evaluation results with respect~to the mono-lingual, cross-lingual and bi-lingual scenarios.}

\textbf{Mono-Lingual Scenario.} The result of learning-based approaches in mono-lingual scenario is reported in Table~\ref{table:result-mono-lingual} and the result of rule-based approaches is reported in Table~\ref{table:result-rule-based}. \tosemrevision{Overall, \tool$_{RoBERTa}$ outperforms all the other approaches. In PyPI, \tool$_{RoBERTa}$ achieves a precision of 96.0\% and a recall of 91.7\%. It achieves the best result, with slight advantages compared with \tool$_{BERT}$ and \tool$_{T5}$. It outperforms the state-of-the-art \textsc{SAP} by 12.1\% in precision, and \textsc{MPHunter} by 6.3\% in recall. In NPM, \tool$_{RoBERTa}$ achieves a precision of 98.5\% and a recall of 92.9\%. It outperforms \tool$_{T5}$ with a 1.9\% higher recall and a 0.4\% lower precision, and \tool$_{BERT}$ with a 0.3\% higher precision and 1.1\% higher recall. It outperforms the state-of-the-art \textsc{Amalfi}$_{DT}$ by 5.8\% in precision and 6.9\% in recall.} The two rule-based approaches have low precision and recall. 

\textbf{Cross-Lingual Scenario.} The result of cross-lingual scenario is presented in Table~\ref{table:result-cross-lingual}. 
\tosemrevision{When the model is trained on NPM packages and tested on PyPI packages, \tool$_{RoBERTa}$~obtains a highest precision of 75.1\%. However, \tool$_{T5}$ achieves a more balanced performance with a precision of 65.2\% and a recall of 63.4\%. Although \textsc{Amalfi}${SVM}$ has better recall, \tool$_{T5}$ significantly surpasses it in precision, resulting in an overall F1-Score advantage of 13.3\%. When the model~is~trained~on PyPI packages and tested on NPM packages, \tool$_{RoBERTa}$ obtains a highest precision of 46.5\%, and a recall of 92.7\%. Although \textsc{Amalfi}${SVM}$ has better recall, \tool$_{RoBERTa}$ significantly surpasses it in precision, resulting in a F1-Score advantage of 17.7\%.}
This result potentially owes to our well-abstracted feature set. Notice that \tool in the cross-lingual scenario achieves a lower precision and recall than in the mono-lingual scenario, which indicates that some malicious behaviors might be not common, and thus bi-lingual knowledge fusing is necessary.

\textbf{Bi-Lingual Scenario.} The result of bi-lingual scenario is shown in Table~\ref{table:result-bi-lingual}. 
\tosemrevision{Overall, \tool$_{RoBERTa}$ outperforms all the other approaches in PyPI and NPM packages. For testing PyPI packages using the model trained on mixed packages, \tool$_{BERT}$ obtains a precision of 95.0\% and a recall of 92.0\%, and \tool$_{RoBERTa}$ obtains a precision of 96.1\% and a recall of 90.9\%. They achieve very similar results, with \tool$_{RoBERTa}$ outperforming \tool$_{BERT}$ by 1.1\% in precision, while \tool$_{BERT}$ surpasses \tool$_{RoBERTa}$ by the same margin in recall. Comparing with the state-of-the-art, \tool$_{RoBERTa}$ outperform \textsc{Amalfi}$_{DT}$ by 12.9\% in precision and 9.8\% in recall. For testing NPM packages using the model trained on mixed packages, \tool$_{RoBERTa}$ achieves a precision of 98.9\% and a recall of 93.9\%. It outperforms \tool$_{BERT}$ by 0.2\% in precision and 0.9\% in recall, and surpasses \tool$_{T5}$ by 1.9\% in recall. Comparing with the state-of-the-art, \tool$_{RoBERTa}$ outperforms \textsc{Amalfi}$_{DT}$ by 6.8\% in precision 8.1\% in recall. Although \textsc{Amalfi}$_{SVM}$ outperforms \tool in recall by 0.7\%, it suffers from a very low precision of 28.8\%.}

\textbf{Comparing Bi-lingual with Mono-lingual.} \tosemrevision{Comparing the effectiveness of \tool in bi-lingual and mono-lingual scenarios, we observe that \tool$_{BERT}$, \tool$_{RoBERTa}$, and \tool$_{T5}$ trained on mixed packages outperform their mono-lingual counterparts. For detecting malicious PyPI packages, the bi-lingual model of \tool achieves an average precision increase of 0.2\% and an average recall increase of 0.2\%. For NPM packages, the bi-lingual model achieves an average precision increase of 0.3\% and an average recall increase of 1.1\%. These results indicate that our bi-lingual knowledge fusion is useful.}

\tosemminorrevision{When comparing the performance of \tool$_{BERT}$ and \tool$_{RoBERTa}$, we observe that \tool$_{RoBERTa}$ generally outperforms \tool$_{BERT}$ in most scenarios. This advantage can be attributed to RoBERTa’s use of more advanced training strategies, including dynamic masking, larger batch sizes, and an increased number of training steps, which enable the model to learn more efficiently. However, in cross-lingual and bi-lingual tasks, \tool$_{BERT}$ achieves higher or comparable recall compared to \tool$_{RoBERTa}$. Further analysis reveals that for malicious packages detected by \tool$_{BERT}$ but missed by \tool$_{RoBERTa}$, the sequences tend to be longer. This suggests that \tool$_{RoBERTa}$ may be less effective at capturing long-range dependencies within longer sequences, which are essential for recognizing complex malicious patterns. A potential explanation for this discrepancy lies in the pre-training tasks of the two models: RoBERTa does not include the Next Sentence Prediction (NSP) task during its training, while BERT does. The inclusion of NSP in BERT's training may enhance its ability to comprehend long-context relationships, making it better at detecting certain malicious packages that RoBERTa misses.}


\textbf{Summary.} \tosemrevision{\tool$_{RoBERTa}$ outperforms the state-of-the-art by averagely 10.0\% in precision and 7.4\% in recall in the mono-lingual scenario, and averagely 9.9\% in precision and 8.9\% in recall in the bi-lingual scenario. \tool learns bi-lingual knowledge from two package registries, which contributes to an average increase of 0.3\% in precision and 0.7\% in recall.}



\begin{table}[!t]
    \centering
    \footnotesize
    \caption{\tosemrevision{Evaluation Result in Cross-Lingual Scenario}}\label{table:result-cross-lingual}
    \vspace{-10pt}
    \begin{tabular}{m{0.4cm}m{0.34cm}m{0.46cm}m{0.46cm}m{0.46cm}m{0.46cm}m{0.46cm}m{0.46cm}m{0.46cm}m{0.46cm}m{0.46cm}m{0.46cm}m{0.51cm}m{0.51cm}m{0.46cm}m{0.46cm}m{0.46cm}m{0.46cm}}   
        \noalign{\hrule height 1pt}
        \multirow{2}{*}{Train}  & \multirow{2}{*}{Test} & \multicolumn{2}{c}{\shortstack[l]{\tool$_{BERT}$}} & \multicolumn{2}{c}{\tosemrevision{\shortstack[l]{\tool$_{RoBERTa}$}}} & \multicolumn{2}{c}{\tosemrevision{\shortstack[l]{\tool$_{T5}$}}}   & \multicolumn{2}{c}{\shortstack[l]{\textsc{Amalfi}$_{DT}$}} & \multicolumn{2}{c}{\shortstack[l]{\textsc{Amalfi}$_{NB}$}} & \multicolumn{2}{c}{\shortstack[l]{\textsc{Amalfi}$_{SVM}$}}\\
    \cmidrule(lr){3-4}
    \cmidrule(lr){5-6}
    \cmidrule(lr){7-8}
    \cmidrule(lr){9-10}
    \cmidrule(lr){11-12}
    \cmidrule(lr){13-14}
    & & Pre. & Rec. & Pre. & Rec. & Pre. & Rec. & Pre. & Rec. & Pre. & Rec. & Pre. & Rec. \\ 
    \noalign{\hrule height 1pt}
    NPM & PyPI & 72.0\% & 49.1\% & \tosemrevision{75.1\%} & \tosemrevision{44.1\%} & \tosemrevision{65.2\%} & \tosemrevision{63.4\%} & 72.0\% & 18.0\% & 42.0\% & 31.5\% & 38.4\% & 76.0\% \\
    PyPI & NPM & 41.9\% & 93.0\% & \tosemrevision{46.5\%} & \tosemrevision{92.7\%} & \tosemrevision{37.5\%} & \tosemrevision{90.9\%} & 14.4\% & 4.7\% & 48.8\% & 5.2\% & 28.8\% & 94.6\% \\
    \noalign{\hrule height 1pt}
    \end{tabular}
\end{table}

\begin{table}[!t]
    \centering
    \footnotesize
    \caption{\tosemrevision{Evaluation Result in Bi-Lingual Scenario}}\label{table:result-bi-lingual}
    \vspace{-10pt}
    \begin{tabular}{m{0.45cm}m{0.40cm}m{0.46cm}m{0.46cm}m{0.46cm}m{0.46cm}m{0.46cm}m{0.46cm}m{0.46cm}m{0.46cm}m{0.46cm}m{0.46cm}m{0.51cm}m{0.51cm}m{0.46cm}m{0.46cm}}   
        \noalign{\hrule height 1pt}
        \multirow{2}{*}{Train}  & \multirow{2}{*}{Test}  & \multicolumn{2}{c}{\shortstack[l]{\tool$_{BERT}$}} & \multicolumn{2}{c}{\tosemrevision{\shortstack[l]{\tool$_{RoBERTa}$}}} & \multicolumn{2}{c}{\tosemrevision{\shortstack[l]{\tool$_{T5}$}}} & \multicolumn{2}{c}{\shortstack[l]{\textsc{Amalfi}$_{DT}$}} & \multicolumn{2}{c}{\shortstack[l]{\textsc{Amalfi}$_{NB}$}} & \multicolumn{2}{c}{\shortstack[l]{\textsc{Amalfi}$_{SVM}$}} & \multicolumn{2}{c}{\tosemrevision{\textsc{SAP}}}\\
    \cmidrule(lr){3-4}
    \cmidrule(lr){5-6}
    \cmidrule(lr){7-8}
    \cmidrule(lr){9-10}
    \cmidrule(lr){11-12}
    \cmidrule(lr){13-14}
    \cmidrule(lr){15-16}
    & & Pre. & Rec. & Pre. & Rec. & Pre. & Rec. & Pre. & Rec. & Pre. & Rec. & Pre. & Rec. & Pre. & Rec. \\ 
    \noalign{\hrule height 1pt}
    \multirow{2}{*}{Mixed}  & PyPI & 95.0\% & 92.0\% & \tosemrevision{96.1\%} & \tosemrevision{90.9\%} & \tosemrevision{94.6\%} & \tosemrevision{69.6\%} & 83.2\% & 81.1\% & 42.0\% & 35.6\% & 45.1\% & 37.2\% & \tosemrevision{80.2\%} & \tosemrevision{60.2\%} \\
     & NPM & 98.7\% & 93.0\% & \tosemrevision{98.9\%} & \tosemrevision{93.9\%} & \tosemrevision{98.9\%} & \tosemrevision{92.0\%} & 92.1\% & 85.8\% & 93.9\% & 76.6\% & 28.8\% & 94.6\% & \tosemrevision{90.4\%} & \tosemrevision{39.5\%} \\
    \noalign{\hrule height 1pt}
    \end{tabular}
\end{table}
 

\subsection{Efficiency Evaluation (RQ2)}


We measure the time overhead of \tool for training and prediction, including all components described in Section~\ref{approach}. \tool is trained on a machine with an Intel Xeon(R) Silver 4314 CPU at 2.40GHz, 128G of RAM and an NVIDIA GeForce RTX 3090. The same machine is employed~for predictions. \tosemrevision{Table \ref{tab:performance} presents the efficiency results.} To train \tool with packages from a single package registry, it takes around \todo{14.1} hours for PyPI packages and \todo{25.2} hours for NPM packages. To equip \tool with bi-lingual knowledge using the mixed packages, the training time is proportional to the scale of the dataset, which amounts to \todo{39.3} hours. Moreover, the average prediction time for a package is similar for \tosemrevision{BERT, RoBERTa, or T5} trained on PyPI, NPM and mixed packages, which is respectively \todo{11.552}, \todo{10.030} and \todo{10.527}~seconds. 

\begin{table}[!t]
    \footnotesize
    \caption{\tosemrevision{Efficiency Evaluation Results}}\label{tab:performance}
    \vspace{-10pt}
    \begin{tabular}{cccccc}
        \noalign{\hrule height 1pt}
        \multirow{2}{*}{Training Data} & \multirow{2}{*}{Training Time (h)} & \multicolumn{3}{c}{Average Prediction Time (s)} \\
        \cmidrule(lr){3-6}
        & & Feature Extractor & Behavior Sequence Generator & Maliciousness Classifier & Total \\
        \hline
        \tosemrevision{PyPI} & \tosemrevision{14.1} & \tosemrevision{3.737} & \tosemrevision{7.807} & \tosemrevision{0.008} & \tosemrevision{11.552} \\
        \tosemrevision{NPM} & \tosemrevision{25.2} & \tosemrevision{2.412} & \tosemrevision{7.611} & \tosemrevision{0.007} & \tosemrevision{10.030} \\
        \tosemrevision{Mixed} & \tosemrevision{39.3} & \tosemrevision{2.856} & \tosemrevision{7.664} & \tosemrevision{0.007} & \tosemrevision{10.527} \\
        \noalign{\hrule height 1pt}
    \end{tabular}
\end{table}

\textbf{\textit{Summary.}} \tool takes less than two days to train the model; and \tool takes an average of \todo{10.5} seconds to predict whether a package is malicious, which is acceptable.

\begin{figure*}[!t]
    \centering
    \begin{subfigure}[b]{0.49\linewidth}
        \includegraphics[width=\linewidth]{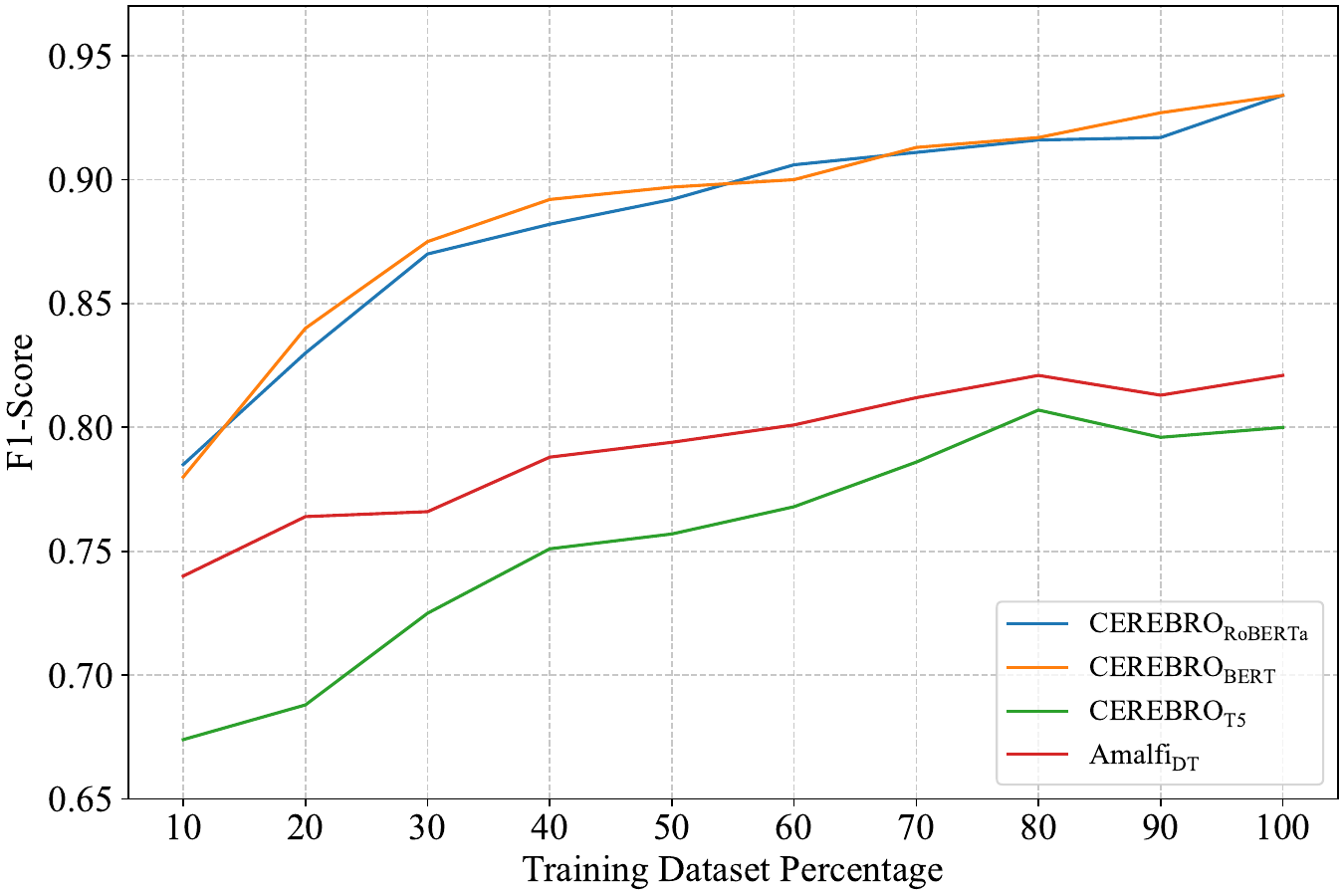}
        \vspace*{-5pt}
        \caption{\tosemminorrevision{PyPI Packages as the Testing Dataset}}
        \label{fig:mixed_to_pypi}
    \end{subfigure}
    \hfill 
    \begin{subfigure}[b]{0.49\linewidth}
        \includegraphics[width=\linewidth]{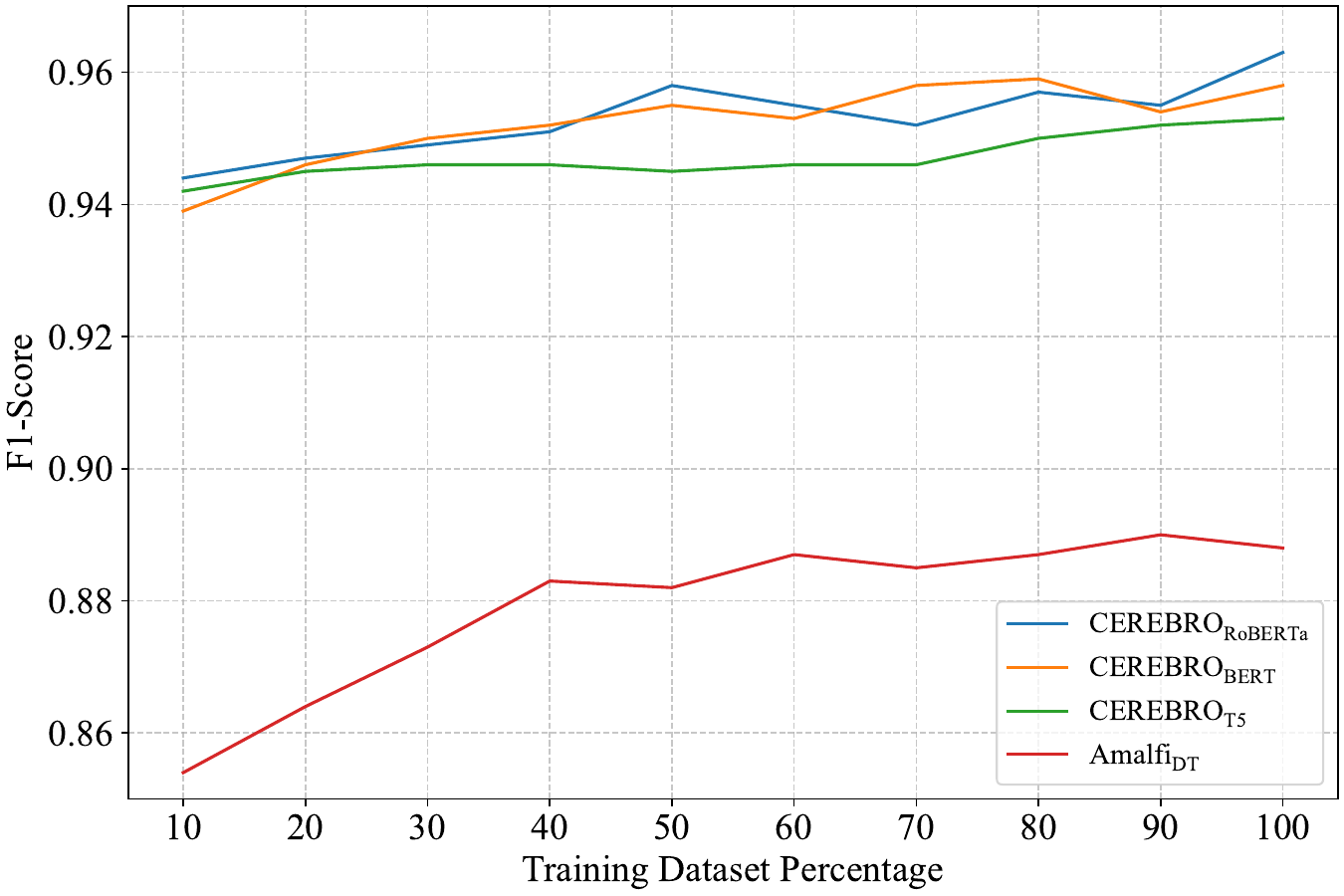}
        \vspace*{-5pt}
        \caption{\tosemminorrevision{NPM Packages as the Testing Dataset}}
        \label{fig:mixed_to_npm}
    \end{subfigure}
    \vspace{-10pt}
    \caption{\tosemminorrevision{Effectiveness of \tool$_{BERT}$, \tool$_{RoBERTa}$, \tool$_{T5}$ and \textsc{Amalfi}$_{DT}$ in Different Scales of Dataset}}
    \label{fig:dataset_scale}
\end{figure*}

\subsection{\tosemrevision{Dataset Scale Evaluation (RQ3)}}


\tosemrevision{Figure~\ref{fig:dataset_scale} presents the results of our dataset scale evaluation. Overall, \tool$_{BERT}$ and \tool$_{RoBERTa}$ consistently outperform the state-of-the-art by a significant margin in both PyPI and NPM packages, even with limited dataset scales (i.e., 10\%, 20\%). In Figure~\ref{fig:dataset_scale}(a), \tool$_{BERT}$ and \tool$_{RoBERTa}$ exhibit similar F1-Scores as the dataset scale changes. \tool$_{RoBERTa}$ outperforms \textsc{Amalfi}${DT}$ by an F1-Score of 0.045 at the 10\% scale. As the dataset scale increases, \tool$_{RoBERTa}$ further widens the gap, surpassing \textsc{Amalfi}${DT}$ with an F1-Score of 0.113 at the 100\% scale. In Figure~\ref{fig:dataset_scale}(b), \tool$_{BERT}$, \tool$_{RoBERTa}$, and \tool$_{T5}$ exhibit similar F1-Scores as the dataset scale changes. \tool$_{RoBERTa}$ outperforms \textsc{Amalfi}${DT}$ by an F1-Score of 0.09 at the 10\% scale. As the dataset scale increases, \tool$_{RoBERTa}$ further keeps the gap, achieving an average F1-Score difference of 0.072 across remaining dataset scales.} 
\tosemminorrevision{We also observe that as the training dataset increases, the F1-score improvement for NPM packages is relatively smaller compared to PyPI packages. This is primarily due to the differing levels of package diversity between the two ecosystems. As shown in Fig. \ref{fig:malware_category}, a larger proportion of NPM packages focus on information stealing. Consequently, models trained on NPM packages quickly learn and generalize malicious patterns from smaller datasets than those for PyPI, resulting in better performance at lower data percentages and smaller incremental gains as the dataset scale increases.}

\tosemrevision{\textbf{Summary.} The effectiveness of our approach maintains an advantage over state-of-the-art across different dataset scales. \tool$_{RoBERTa}$ achieves an F1-Score of \tosemrevision{0.785} in PyPI and \tosemrevision{0.944} in NPM even when the dataset scale is reduced by 90\%, indicating the robustness of \tool in situations with scarce data availability.}

\subsection{Ablation Study (RQ4)}


We train the three ablated versions of \tool in the bi-lingual scenario. As shown in Table~\ref{table:result-ablation}, the ablated version is less effective than the original version.
\tosemrevision{Specifically, the average precision and recall for \tool$_{BERT}$, \tool$_{RoBERTa}$ and \tool$_{T5}$ decrease 1.6\% and 0.6\%, increase 1.6\% and decrease 3.7\%, decrease 1.4\% and 1.9\% for \tool w/o Seq in mono-lingual, cross-lingual and bi-lingual scenarios, respectively.}
\tosemrevision{Similarly, the average precision and recall decrease 1.4\% and 1.2\%, decrease 2.0\% and 33.0\%, decrease 0.5\% and 0.7\% for \tool w/o Text in mono-lingual, cross-lingual and bi-lingual scenarios, respectively.}
\tosemrevision{Further, the average precision and recall decrease 3.0\% and 0.2\%, 27.1\% and 61.4\%, 4.7\% and 3.2\% for \tool w/ DT in mono-lingual, cross-lingual and bi-lingual scenarios, respectively.}

\begin{table}[!t]
    \centering
    \tiny
    \caption{\tosemrevision{Evaluation Result on Ablated Versions of \tool (The bold numbers indicate the decreases or increases compared to the non-ablated versions)}}\label{table:result-ablation}
    \vspace{-10pt}
    \begin{tabular}{m{0.30cm}m{0.30cm}m{0.55cm}m{0.55cm}m{0.55cm}m{0.55cm}m{0.55cm}m{0.55cm}m{0.55cm}m{0.55cm}m{0.55cm}m{0.55cm}m{0.55cm}m{0.55cm}m{0.55cm}m{0.55cm}}   
        \noalign{\hrule height 1pt}
        \multirow{3}{*}{Train} & \multirow{3}{*}{Test} & \multicolumn{6}{c}{\shortstack[l]{\textsc{\tool} w/o Seq}} & \multicolumn{6}{c}{\shortstack[l]{\textsc{\tool} w/o Text}} & \multicolumn{2}{c}{\shortstack[l]{\tool w/ DT}}\\
    \cmidrule(lr){3-8}
    \cmidrule(lr){9-14}
    \cmidrule(lr){15-16}
    & & \multicolumn{2}{c}{\textsc{BERT}} & \multicolumn{2}{c}{\tosemrevision{\textsc{RoBERTa}}} & \multicolumn{2}{c}{\tosemrevision{\textsc{T5}}} &  \multicolumn{2}{c}{\textsc{BERT}} & \multicolumn{2}{c}{\tosemrevision{\textsc{RoBERTa}}} & \multicolumn{2}{c}{\tosemrevision{\textsc{T5}}} & \multirow{2}{*}{Pre.} & \multirow{2}{*}{Rec.} \\
    \cmidrule(lr){3-4}
    \cmidrule(lr){5-6}
    \cmidrule(lr){7-8}
    \cmidrule(lr){9-10}
    \cmidrule(lr){11-12}
    \cmidrule(lr){13-14}
    & & Pre. & Rec. & Pre. & Rec. & Pre. & Rec. & Pre. & Rec. & Pre. & Rec. & Pre. & Rec. & & \\ \hline
    \tosemrevision{PyPI} & \tosemrevision{PyPI} & 91.3\% \textbf{\newline $\downarrow$ 4.5\%} & 87.1\% \textbf{\newline $\downarrow$ 2.3\%} & \tosemrevision{92.1\% \textbf{\newline $\downarrow$ 3.9\%}} & \tosemrevision{86.2\% \textbf{\newline $\downarrow$ 5.5\%}} & \tosemrevision{92.1\% \textbf{\newline $\downarrow$ 1.3\%}} & \tosemrevision{76.1\% \textbf{\newline $\uparrow$ 5.2\%}} & 92.6\% \textbf{\newline $\downarrow$ 3.2\%} & 85.0\% \textbf{\newline $\downarrow$ 4.4\%} & \tosemrevision{90.6\% \textbf{\newline $\downarrow$ 5.4\%}} & \tosemrevision{89.6\% \textbf{\newline $\downarrow$ 2.1\%}} & \tosemrevision{94.1\% \textbf{\newline $\uparrow$ 0.7\%}} & \tosemrevision{76.8\% \textbf{\newline $\uparrow$ 5.9\%}} & \tosemrevision{91.4\% \textbf{\newline $\downarrow$ 4.4\%}} & \tosemrevision{89.4\% \textbf{\newline $\downarrow$ 0.0\%}} \\
    \tosemrevision{NPM} & \tosemrevision{NPM} & 98.0\% \textbf{\newline $\downarrow$ 0.2\%} & 92.2\% \textbf{\newline $\uparrow$ 0.4\%} &\tosemrevision{98.9\% \textbf{\newline $\uparrow$ 0.4\%}} & \tosemrevision{92.1\% \textbf{\newline $\downarrow$ 0.8\%}} & \tosemrevision{98.8\% \textbf{\newline $\downarrow$ 0.1\%}} & \tosemrevision{90.4\% \textbf{\newline $\downarrow$ 0.6\%}} & 98.1\% \textbf{\newline $\downarrow$ 0.1\%} & 90.4\% \textbf{\newline $\downarrow$ 1.4\%} & \tosemrevision{97.7\% \textbf{\newline $\downarrow$ 0.8\%}} & \tosemrevision{90.4\% \textbf{\newline $\downarrow$ 2.5\%}} & \tosemrevision{99.1\% \textbf{\newline $\uparrow$ 0.2\%}} & \tosemrevision{88.3\% \textbf{\newline $\downarrow$ 2.7\%}} &  \tosemrevision{96.6\% \textbf{\newline $\downarrow$ 1.6\%}} & \tosemrevision{92.2\% \textbf{\newline $\uparrow$ 0.4\%}} \\
    \hline 
    \tosemrevision{NPM} & \tosemrevision{PyPI} & 77.3\% \textbf{\newline $\uparrow$ 5.3\%} & 36.0\% \textbf{\newline $\downarrow$ 13.1\%} & \tosemrevision{76.7\% \textbf{\newline $\uparrow$ 1.6\%}} & \tosemrevision{42.1\% \textbf{\newline $\downarrow$ 2.0\%}} & \tosemrevision{64.7\% \textbf{\newline $\downarrow$ 0.5\%}} & \tosemrevision{67.6\% \textbf{\newline $\uparrow$ 4.2\%}} & 93.7\% \textbf{\newline $\uparrow$ 21.7\%} & 18.5\% \textbf{\newline $\downarrow$ 30.6\%} & \tosemrevision{64.2\% \textbf{\newline $\downarrow$ 10.9\%}} & \tosemrevision{2.0\% \textbf{\newline $\downarrow$ 42.1\%}} & \tosemrevision{88.0\% \textbf{\newline $\uparrow$ 22.8\%}} & \tosemrevision{8.4\% \textbf{\newline $\downarrow$ 55.0\%}} & \tosemrevision{49.5\% \textbf{\newline $\downarrow$ 22.5\%}} & \tosemrevision{11.0\% \textbf{\newline $\downarrow$ 38.1\%}} \\
    \tosemrevision{PyPI} & \tosemrevision{NPM}  & 49.2\% \textbf{\newline $\uparrow$ 7.3\%} & 92.2\% \textbf{\newline $\downarrow$ 0.8\%} & \tosemrevision{38.3\% \textbf{\newline $\downarrow$ 8.2\%}} & \tosemrevision{87.2\% \textbf{\newline $\downarrow$ 5.5\%}} & \tosemrevision{41.3\% \textbf{\newline $\uparrow$ 3.8\%}} & \tosemrevision{86.0\% \textbf{\newline $\downarrow$ 4.9\%}} & 41.8\% \textbf{\newline $\downarrow$ 0.1\%} & 94.0\% \textbf{\newline $\uparrow$ 1.0\%} & \tosemrevision{28.0\% \textbf{\newline $\downarrow$ 18.5\%}} & \tosemrevision{98.9\% \textbf{\newline $\uparrow$ 6.2\%}} & \tosemrevision{10.6\% \textbf{\newline $\downarrow$ 26.9\%}} & \tosemrevision{13.6\% \textbf{\newline $\downarrow$ 77.3\%}} &  \tosemrevision{10.2\% \textbf{\newline $\downarrow$ 31.7\%}} & \tosemrevision{8.4\% \textbf{\newline $\downarrow$ 84.6\%}} \\
    \hline
    \multirow{2}{*}{Mixed} & PyPI & 92.3\% \textbf{\newline $\downarrow$ 2.7\%} & 86.2\% \textbf{\newline $\downarrow$ 5.8\%} & \tosemrevision{93.4\% \textbf{\newline $\downarrow$ 2.7\%}} & \tosemrevision{88.3\% \textbf{\newline $\downarrow$ 2.6\%}} & \tosemrevision{92.6\% \textbf{\newline $\downarrow$ 2.0\%}} & \tosemrevision{71.3\% \textbf{\newline $\uparrow$ 1.7\%}} & 95.5\% \textbf{\newline $\uparrow$ 0.5\%} & 86.0\% \textbf{\newline $\downarrow$ 6.0\%} & \tosemrevision{94.7\% \textbf{\newline $\downarrow$ 1.4\%}} & \tosemrevision{87.4\% \textbf{\newline $\downarrow$ 3.5\%}} & \tosemrevision{92.9\% \textbf{\newline $\downarrow$ 1.7\%}} & \tosemrevision{82.0\% \textbf{\newline $\uparrow$ 12.4\%}} &  90.5\% \textbf{\newline $\downarrow$ 4.5\%} & 86.6\% \textbf{\newline $\downarrow$ 5.4\%} \\
     & NPM  & 98.2\% \textbf{\newline $\downarrow$ 0.5\%} & 91.8\% \textbf{\newline $\downarrow$ 1.2\%} & \tosemrevision{98.6\% \textbf{\newline $\downarrow$ 0.3\%}} & \tosemrevision{91.7\% \textbf{\newline $\downarrow$ 2.2\%}} & \tosemrevision{98.9\% \textbf{\newline $\downarrow$ 0.0\%}} & \tosemrevision{90.9\% \textbf{\newline $\downarrow$ 1.1\%}} & 98.7\% \textbf{\newline $\downarrow$ 0.0\%} & 91.1\% \textbf{\newline $\downarrow$ 1.9\%} & \tosemrevision{98.3\% \textbf{\newline $\downarrow$ 0.6\%}} & \tosemrevision{91.1\% \textbf{\newline $\downarrow$ 2.8\%}} & \tosemrevision{98.8\% \textbf{\newline $\downarrow$ 0.1\%}} & \tosemrevision{89.5\% \textbf{\newline $\downarrow$ 2.5\%}} & 93.9\% \textbf{\newline $\downarrow$ 4.8\%} & 92.0\% \textbf{\newline $\downarrow$ 1.0\%} \\
    \noalign{\hrule height 1pt}
    \end{tabular}
\end{table}

\begin{table*}[!t]
    \centering
    \footnotesize
    \caption{\new{Result of Real-World Malicious Package Detection \todel{Over \todo{Six} Weeks}}}\label{table:monitor-stats}
    \vspace{-10pt}
    \begin{tabular}{m{1.1cm}m{1.2cm}m{2.2cm}m{1.2cm}m{1.1cm}m{1.1cm}m{1.5cm}m{1.5cm}}
        \noalign{\hrule height 1pt}
        Package Registry & Time Period & Newly Published Package Versions & Flagged by \tool & False Positives & True Positives & Thank Letters & Removed Before Report \\
        \hline
        \multirow{7}{*}{PyPI}
        & March & 104,512 & 750 & 521 & 229 & 173 & 56 \\
        & April & 101,415 & 661 & 560 & 101 & 0 & 101 \\
        & May & 78,528 & 422 & 324 & 98 & 30 & 68 \\
        & June & 102,088 & 768 & 595 & 173 & 0 & 173 \\
        & July & 101,600 & 720 & 670 & 50 & 10 & 40 \\
        & August & 56,561 & 298 & 283 & 15 & 14 & 1 \\
        & September & 25,193 & 101 & 88 & 13 & 4 & 9 \\
        & October & 29,596 & 16 & 12 & 4 & 1 & 3 \\
        \cline{2-8}
        & Total & 599,493 & 3,746 & 3,053 & 683 & 232 & 451 \\
        \hline
        \multirow{6}{*}{NPM}
        & April & 130,697 & 325 & 178 & 147 & 147 & 0 \\
        & May & 39,094 & 145 & 88 & 57 & 35 & 22 \\
        & June & 28,764 & 256 & 145 & 111 & 0 & 111 \\
        & July & 34,626 & 244 & 155 & 89 & 44 & 45 \\
        & August & 33,013 & 514 & 314 & 200 & 131 & 69 \\
        & September & 27,350 & 374 & 254 & 120 & 91 & 29 \\
        & October & 30,601 & 372 & 297 & 75 & 27 & 48 \\
        \cline{2-8}
        & Total & 324,145 & 2,230 & 1431 & 799 & 475 & 324 \\
        \hline
        \noalign{\hrule height 1pt}
    \end{tabular}
\end{table*}

\textbf{\textit{Summary.}} Generating behavior sequence, transforming behavior sequence into textual description, and feeding textual description into a classifier fine-tuned from a \tosemrevision{pre-trained language model} are all effective in detecting malicious packages. \tosemrevision{Averagely, the ablated versions of \tool decrease 4.5\% in precision and 11.8\% in recall.}

%

\subsection{Real-World Usefulness Evaluation (RQ5)}

To evaluate the practical usefulness of \tool in analyzing newly published package versions in real-world registries,~we design a monitoring system that first monitors and crawls newly published package versions from PyPI and NPM and then runs \tool against the crawled package versions. Our monitoring system has lasted over \new{\todo{8} months for PyPI and \todo{7} months for NPM}\todel{\todo{six} weeks}.


\textbf{Overall Usefulness Result.} \new{In total,} \tool scans \new{923,638} newly published package versions, and flags \new{3,746} PyPI package versions and \new{2,230} NPM package versions as potentially malicious, as reported in Table \ref{table:monitor-stats}.~We~first manually validate each flagged package version, and confirm \new{683} PyPI package versions and \new{799} NPM package versions as true positives. Then, following responsible disclosure process, we promptly report these package versions to administrators of the official PyPI and NPM teams. At the time of reporting, \new{451} PyPI package versions and \new{324} NPM package versions have already been removed. The PyPI and NPM teams confirm the remaining \new{232} PyPI package versions and \new{475} NPM package versions as malicious, and remove all of them. Thus, we receive~\new{232} and \new{457} thank letters from PyPI and NPM.


\textbf{False Positive Analysis.} As shown in Table \ref{table:monitor-stats}, the false positive rate is \new{81.5\%}\todel{\todo{73.9\%}} for PyPI and \new{64.2\%}\todel{\todo{58.3\%}} for NPM, which is actually high. It is worth mentioning that it takes one~of~our authors around one hour to manually analyze and confirm the flagged package versions every day, which is still acceptable. After analyzing all the false positives, we identify~four major types of benign packages that are flagged as malicious,~as~their behavior sequences are similar to malicious ones. 

First, some benign packages encapsulate RESTful APIs, which enable data transmission~to~remote servers. Their behavior sequences consist of features about \textit{Information Reading} and \textit{Data Transmission} (e.g., R1, D1, D2 and D3). Second, some benign packages encapsulate CLI commands into Python/JavaScript methods where their behavior sequences resemble those malicious packages in using features about \textit{Payload Execution}. Third, some benign packages behave similarly to Trojans, but their payloads serve benign purposes. Normally, their behavior sequences resemble Trojans by starting with D1 and D3, then undertaking intermediate actions such as R1 and R4, and ending with P2 and P3. Last, some benign packages consist of a single meta utility module that incorporates features ranging from \textit{Information Reading}, \textit{Data Transmission}, \textit{Encoding} to \textit{Payload Execution}. It causes \tool to recognize parts of behavior~sequences as similar to those of malicious packages.

\new{\textbf{False Negative Analysis.} To measure false negatives of \tool, we reached out to package registries. However, PyPI replied that they did not maintain a malicious package list and NPM refused to share the list. Therefore, we try to use state-of-the-art detectors to identify true positives. Specifically, we employ \textsc{Amalfi}~\cite{sejfia2022practical} to identify potentially malicious packages~during the same monitoring time period. It flagged \todo{41,434} PyPI package versions and \todo{1,930} NPM package versions as potentially malicious. Then, we undertake a manual inspection process with a sampling rate of 95\% confidence level and 5\% error margin, which results in \todo{381} PyPI packages and \todo{164} NPM packages. After manual validation, we identify \todo{9} malicious PyPI packages and \todo{45} malicious NPM packages. Remarkably, only \todo{2} PyPI packages and \todo{5} NPM packages went undetected by \tool. It indicates that the false negative rate of \tool is relatively low in real-world detection.}



\begin{table}[t]
    \footnotesize
    \centering
    \caption{\new{Result of Incremental Learning}}\label{table:result-after-retrain}
    \vspace{-10pt}
    \begin{tabular}{m{2.6cm}m{2.6cm}m{1.6cm}m{1.0cm}}
        \toprule
        Added Train (\#) & Time Period  & New Test (\#)  & Pre.\\\hline
        None &  - & \multirow{5}{*}{PyPI (1,135)} & 7.2\% \\
        PyPI (750) & March &  & 26.5\% \\
        Mixed (1,736) & March-April &  & 36.5\% \\
        Mixed (2,303) & March-May &  & 39.0\% \\
        Mixed (3,327) & March-June &  & 58.2\% \\
        \cline{1-4}
        None & -- & \multirow{5}{*}{NPM (1,504)} & 32.2\%  \\
        PyPI (750) & March &  & 76.3\% \\
        Mixed (1,736) & March-April &  & 80.3\% \\
        Mixed (2,303) & March-May &  & 80.2\% \\
        Mixed (3,327) & March-June &  & 80.7\% \\
        \bottomrule
    \end{tabular}
\end{table}

\textbf{Incremental Learning Result.} As we are continuously confirming and accumulating true positives and false positives of malicious packages based on our monitoring system, one potential way to reduce false positives is to adopt incremental learning on such true positives and false positives. \new{To this~end, we add the true positives and false positives confirmed\tosemrm{during the first ten days and the \tosem{first} three months} during our monitoring \tosem{from March to June} as two incremental datasets for retraining.} \tosem{In particular, we add 750 packages from March, 1,736 packages from March to April, 2,303 packages from March to May and 3,327 packages from March to June, as reported in the third to sixth row, and eight to eleven row under the column of \textit{Added Train} in Table~\ref{table:result-after-retrain}. Note that the added training dataset for March contains only PyPI packages because monitoring NPM packages starts from April.} \new{Then, we retrain four models of \tool by augmenting the original mixed training dataset with the new mixed PyPI and NPM packages.} Finally, we select true positives and false positives confirmed during our monitoring from July to October as the new testing dataset. In particular, there are \todo{1,135} PyPI packages \new{and \todo{1,504} NPM packages } as reported in the third column in Table \ref{table:result-after-retrain}. 

Overall, by incorporating new malicious and benign packages into the training dataset, the precision of \tool~improves significantly. Specifically, when \tool is retrained with new mixed packages, the precision of detecting malicious PyPI packages reaches \new{58.2\%}, while the precision of detecting malicious NPM packages reaches \new{80.7\%}, achieving a precision increase of \new{51.0\%}\todel{\todo{22.7\%}} for PyPI packages and \todel{\todo{28.6\%}}\new{48.5\%} for NPM packages in total.

\begin{figure*}[!t]
    \centering
    \begin{subfigure}[t]{0.45\linewidth} 
        \centering
        \includegraphics[width=\linewidth]{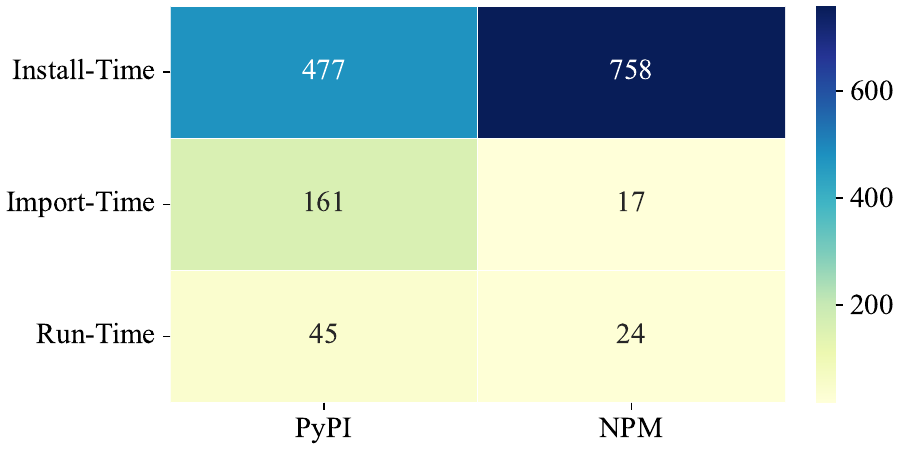}
        \caption{\tosemminorrevision{Triggering Scenarios of Malicious Packages}}
        \label{fig:trigger_type}
    \end{subfigure}
    \hfill 
    \begin{subfigure}[t]{0.45\linewidth} 
        \centering
        \includegraphics[width=\linewidth]{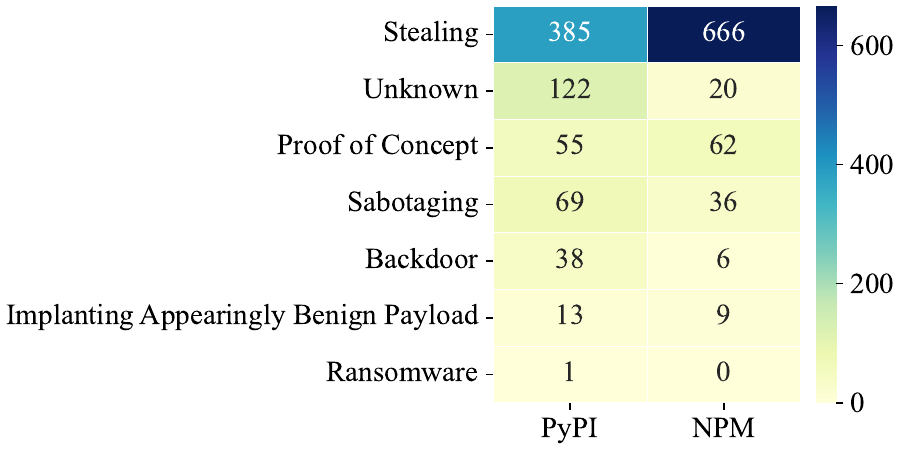}
        \caption{\tosemminorrevision{Malicious Intentions of Malicious Packages}}
        \label{fig:malware_category}
    \end{subfigure}
    \vspace{-7pt}  
    \caption{\tosemminorrevision{Triggering Scenarios and Malicious Intentions of New Malicious Packages}} 
\end{figure*}


\textbf{Malicious Package Characteristics.} For each of the \new{683} PyPI and \new{799} NPM malicious package versions which are previously unknown, we manually look into their malicious behavior to understand~their~characteristics. On the one hand, we determine their triggering~scenarios, and report the result \tosemrevision{in Figure~\ref{fig:trigger_type}}. We can observe that \new{477 (69.8\%)} of the malicious PyPI package versions and \new{758 (94.9\%)} of the malicious NPM package versions trigger their malicious behavior at install-time. This is reasonable because install-time execution can achieve the highest probability of successfully carrying out attacks. \new{Moreover, a small number of malicious PyPI package versions opt to trigger malicious~behavior other than install-time, accounting~for~\new{161~(23.6\%)} at import-time and 45 (6.6\%) at run-time. Noticeably, there are only \new{17 (2.1\%)} of the malicious NPM package versions that adopt import-time triggering, and \new{24 (\new{3.0\%})} malicious NPM package versions adopt run-time triggering.}

On the other hand, we explore the intentions of malicious package versions by executing them in a sandbox, and present the result in \tosemrevision{Figure~\ref{fig:malware_category}}. Generally, the malicious intentions~include stealing (e.g., personal data or credentials), sabotaging (e.g., shutting down firewalls), proof~of~concept~(which is harmless but to demonstrate something malicious can be done), backdoor (i.e., enabling remote-controlled operations for an attack to exploit an infected machine), implanting appearingly benign payload (e.g., the benign payload might change to be malicious via morphing Trojan), and ransomware. Besides, we fail to execute the payload for \new{142} of the malicious package versions due to non-existent URLs or run-time exceptions, and thus we label their malicious intentions as unknown.

\textbf{\textit{Summary.}} \tool detects \new{683} and \new{799} previously unknown malicious PyPI and NPM package versions. All of them have been removed by the official PyPI and NPM teams, and we also receive \new{707} thank letters from them. Further, by incorporating new malicious and benign packages, \tool learns new knowledge to improve its precision. Besides, most (\new{83.3\%}) of these new malicious package versions execute the malicious behavior at install-time; and most (\new{70.9\%}) of them attempt to steal sensitive information.




\subsection{Threats and Limitations}

\textbf{Threats.}\todel{The threats to validity are subjected to three aspects. } First,~as the campaign against malware continues, new strains of malicious packages emerge daily. This implies that the original dataset of malicious packages used in our evaluation may not encompass the full spectrum of new malicious packages, hindering \tool's effectiveness. However, we have demonstrated the effectiveness of \tool in learning from newly published packages and adapting to detect emerging threats. Second, while we thoroughly examine~the~potentially malicious package versions identified by \tool, we acknowledge that~the~vast~workload makes it impractical to validate every potentially benign package version. Therefore, we only systematically report the result of precision but not the result of recall. Third, \tool can generate false positives. To address this, we conduct manual validation on identified package versions before reporting them to PyPI and NPM teams. While this process can also introduce false positives, the fact that the reported packages are subsequently removed indicates the validity of our validation. 
\tosemrevision{Fourth, as pre-trained language models continue to advance rapidly, we expect to see new and more powerful models. Nevertheless, the choice of language models is orthogonal to our approach. We plan to test the effectiveness of additional language models on this task.}
\tosemrevision{Fifth, the varying behavior sequences between malicious and benign packages may reduce the accuracy of \tool in cross-lingual scenarios. Nevertheless, the cross-lingual setting represents an extreme case in real-world where there are no malicious samples within the same ecosystem.}


\new{\textbf{Limitations.} \tosem{First, our method is designed for PyPI and NPM packages. However, our method can be easily generalized into other ecosystems by implementing the feature extractor and behavior sequence generator for a new ecosystem. The feature set is language-agnostic. By leveraging the existing AST parser and call graph generator of the new ecosystem, we believe that the integration can be straightforward.} Second, large packages may have a large size of the behavior sequence which can exceed the 512 token limit of BERT. However, based on our analysis of packages in Table \ref{table:monitor-stats}, we identify a relatively modest proportion (\todo{13.6\%}) of packages that exceed the limit. We plan to split the behavior sequence into multiple segments to circumvent the limit while maintaining their sequential integrity.}
\tosemrevision{Third, the proposed approach encounters higher false positives in real-world setting. We foresee several challenges and opportunities that could help reduce false positive rates. On the one hand, we can leverage ensemble methods. Utilizing a combination of different detection tools could be advantageous. An ensemble approach, where decisions are based on the consensus among multiple tools, might improve accuracy and reduce false positive rates. On the other hand, we can utilize dynamic analysis during our detection. Incorporating dynamic analysis into the detection process could provide deeper insights into the behavior of packages by capturing detailed runtime behaviors, potentially reducing false positives by confirming suspicious activities through actual execution rather than static analsysis. Nonetheless, with a large number of new packages being released daily, it remains an open question how to leverage the benefits of dynamic analysis while mitigating its computational and time costs.}










\section{Related Work}


\textbf{Malicious Package Detection.} Recently, there has been an increasing number of OSS supply chain attacks \tosemminorrevision{that} inject malicious code into packages, with package registries such as NPM and PyPI being the largest targets~\cite{pypinpm, hundredsnpm}. Ohm et al.~\cite{ohm2020backstabber} were the first to systematically investigate malicious packages. They collected 174 real-world malicious packages from NPM, PyPI and RubyGems. Using this dataset, they manually analyzed how malicious behavior was injected (e.g., via typosquatting) and triggered (e.g., on installation), what the objective (e.g., data exfiltration) of malicious behavior was, and whether obfuscation was employed. \new{Similarly, Guo et al.~\cite{guo2023empirical} conducted an empirical study to understand the characteristics of the malicious code lifecycle in the PyPI ecosystem.} Various approaches have also been proposed to detect malicious packages, mostly in NPM and PyPI.

In the NPM ecosystem, Zahan et al.~\cite{zahan2022weak} identified six~signals of security weakness in NPM supply chain, and proposed a rule-based malicious package detector \tosemminorrevision{that} searches for keywords in install scripts. Garrett et al.~\cite{garrett2019detecting} selected~features about whether libraries that access the network, file~system and operating system processes are used, whether code~is~evaluated at runtime, and whether new files, new dependencies~and new hookup script entries are present, and adopted clustering to build a benign behavior model which is used to detect malicious NPM packages. Differently, Ohm et al.~\cite{ohm2020supporting}~adopted AST (abstract syntax tree)-level clustering to establish a malicious behavior model. Instead of using unsupervised~methods, Fass et al. \cite{fass2018jast, fass2019jstap} built a random forest classifier based~on the frequency of specific patterns extracted~from AST, CFG (control flow graph) and PDG (program dependency graph) of benign and malicious JavaScript samples. This approach~is specifically designed for obfuscated JavaScript programs, and thus could be ineffective for non-obfuscated ones. Ohm~et~al. \cite{ohm2022feasibility} extended the feature set of Garrett et al.'s work \cite{garrett2019detecting} by including features about package metadata and obfuscation. 
Similarly, Sejfia et al.~\cite{sejfia2022practical} proposed \textsc{Amalfi}, which uses an extended feature set from Garrett et al.'s work \cite{garrett2019detecting} to train classifiers. \todo{These approaches capture malicious behavior as discrete features, hindering their accuracy in detecting malicious packages. Differently, we model malicious behavior in a sequential manner, which can model the maliciousness more precisely.} Notice that Ferreira et al.~\cite{ferreira2021containing} proposed a lightweight permission system to protect applications from malicious package updates at runtime. Recently, Huang et al. \cite{huang2024spiderscan} proposes SpiderScan to identify malicious NPM packages based on graph-based behavior modeling and matching.

In the PyPI ecosystem, Vu et al.~\cite{vu2020typosquatting} proposed a rule-based approach to detect malicious PyPI packages that are spread by typosquatting and combosquatting attacks. The rule checks whether the package has the name similar to a Python standard library or a package with known source code repository. Besides, there are several other rule-based detectors, e.g., Bandit4Mal~\cite{bandit4mal} and Malware Checks~\cite{malwarechecks}. Vu et al.~\cite{vu2023bad}~conducted a comprehensive evaluation of these detectors. 
Instead of relying on rules,~Vu~et al.~\cite{vu2020towards, vu2021lastpymile} identified malicious packages based on discrepancies between the source code repository and distributed artifact of a Python package. Liang et al.~\cite{Liang2021} used both package metadata features and code features to learn an isolated forest model for detecting malicious packages. \new{Recently, Liang et al.~\cite{liang2023needle} utilized clustering techniques to group the PyPI installation scripts so that outliers can be identified.} Sun et al. \cite{sun2024integrating} proposes to integrate deep code behavior features with metadata features to detect malicious PyPI packages. \todo{These approaches share the same limitation with those detectors in NPM because they also model malicious behavior as discrete features.}

Moreover, some approaches support different ecosystems. OSS Detect Backdoor~\cite{Ossdetectbackdoor} is a rule-based detector~that~support 15 ecosystems. Taylor et al.~\cite{taylor2020defending} leveraged package~name similarity and package popularity to detect malicious packages caused by typosquatting in NPM, PyPI and RubyGems. \tosemrevision{Ladisa et al.~\cite{ladisa2023feasibility} utilized language-independent feature (e.g. number of URLs, entropy of strings in code files, etc.) to detect malicious packages in NPM and PyPI. However, these approaches are not based on high-level semantics of source code, which can produce high false positives.} Different from the simple rule in Taylor et al. and Ladisa et al., Duan~et al.~\cite{duan2020towards} derived~five~metadata analysis rules, four static analysis rules and four dynamic analysis rules to detect malicious packages in NPM, PyPI and RubyGems. \todo{However, it is heavyweight as it relies on program analysis.}

Squatting attacks are also common in other domains, e.g., domain name system~\cite{szurdi2014long, agten2015seven, khan2015every, kintis2017hiding, tahir2018s}, container registry~\cite{liu2022exploring} and mobile app~\cite{hu2020mobile}. Malware detection has been widely~studied~\cite{ye2017survey,guo2023empirical}\todel{, but mostly for programs in the binary form}. OSS opens up new opportunities for malware detection due to the availability of source code.




\textbf{Security Threats in OSS Supply Chain.} Apart from malicious packages, there exist various security threats in OSS supply chain. Several studies \cite{cappos2008look, bagmar2021know, zahan2022weak, kaplan2021survey, gu2022investigating} have been conducted to investigate security threats in package registries such as NPM and PyPI. Ladisa et al.~\cite{ladisa2022taxonomy} reported a taxonomy of attacks on all OSS supply chain stages~from code contributions to package distribution. They~also assessed the safeguards against OSS supply chain attacks. Koishybayev et al.~\cite{koishybayev2022characterizing} and Gu et al.~\cite{Gu2023} characterized security threats in continuous integration workflows that produce OSS.

Enck and Williams \cite{enck2022top} summarized five challenges in OSS supply chain security, with the~top one being updating vulnerable dependencies. Specifically, the vulnerability impact analysis along the supply chain has been widely explored in various ecosystems, e.g., NPM~\cite{zerouali2022impact, zimmermann2019small, liu2022demystifying}, Maven~\cite{prana2021out, gkortzis2021software, huang2022characterizing}, PyPI\cite{ruohonen2021large, alfadel2023empirical, prana2021out} and RubyGems~\cite{zerouali2022impact, prana2021out}. To cope with vulnerable dependencies, various software composition analysis tools have been proposed \cite{xu2022insight, ponta2018beyond, liu2022demystifying, huang2022characterizing}.  Besides, several advances have been made on detect maliciousness in different stages of software development. Goyal et al.~\cite{goyal2018identifying} and Gonzalez et al.~\cite{gonzalez2021anomalicious} proposed to detect malicious commits on GitHub. Cao and Dolan-Gavitt~\cite{cao2022fork} tried to identify malware in GitHub forks. Lamb and Zacchiroli~\cite{lamb2021reproducible} and Wheeler \cite{wheeler2005countering} attempted to avoid the injection of maliciousness in compilation by ensuring reproducible builds.

\section{Conclusion and Future Work}

We have proposed and implemented \tool to detect malicious packages in NPM and PyPI. \tool leverages~a~comprehensive feature set that models high-level abstraction~of malicious behavior sequence, enabling bi-lingual knowledge fusing. 
Our extensive evaluation results have demonstrated the effectiveness, efficiency and practical~usefulness of \tool in detecting malicious packages. In the future, we plan to enhance the multi-lingual~capability of \tool by incorporating support for both interpreted and compiled languages. The challenge is to design feature extractors for new languages, and improve the generalizability of our feature set. 
We~also~plan to equip \tool with the capability to pinpoint the specific malicious sequence from the whole behavior sequence, further aiding administrators to go through the manual confirmation promptly.

\section{Data Availability}

\new{The data of our evaluation is available at \todo{https://doi.org/10.5281/zenodo.8277447}.}

\section*{Acknowledgment}

This work was supported by the National Natural Science Foundation of China (Grant No. 62332005, 62372114 and 62402342).

\bibliographystyle{ACM-Reference-Format}
\bibliography{src/reference}

\end{document}